\documentclass[11pt]{article}
\usepackage{graphicx,amsfonts}
\usepackage{biograph,latexsym}
\usepackage{fullpage}
\usepackage{amscd}

\newcommand{\ara}{{ |A_{0}| }}
\newcommand{\ar}{{   A_{0}  }}
\newcommand{\R}{{\mathbb R}}

\newcommand{\N}{{\mathcal N}}
\newcommand{\E}{{\mathrm E}}

\newcommand{\law}{{\mathcal P}}

\newcommand{\mi}{{m(x_{i})}}
\newcommand{\dn}{d_{N}}
\newcommand{\no}{N_0}
\newcommand{\di}{d}

\newcommand{\locj}{{j_{n}( x_{1},\ldots,x_{n}|A)}}

\newcommand{\xk}[1]{{x_{1},\ldots,x_{#1}}}

\newcommand{\dxk}[1]{dx_{1} \cdots dx_{#1}}

\newcommand{\pr}{{\mathbb{P}}}

\newcommand{\ma}{{m_{1}}}
\newcommand{\mb}{{m_{2}}}
\newcommand{\mc}{{m_{3}}}

\newcommand{\mor}{W}
\newcommand{\dis}{B}

\begin{document}
\title{A multiscale maximum entropy moment closure for locally regulated space--time point process models of population dynamics
\thanks{M.R. is grateful for a postgraduate scholarship from the Principal's development fund of the University of Glasgow, an overseas student award granted from the Department of Mathematics, University of Glasgow, and DARPA (Award ID:HR001-05-1-0057)  }
} 
  
     \author{
       Michael Raghib  \and Nicholas A. Hill  \and Ulf Dieckmann}  

\maketitle
\begin{abstract}
The prevalence of structure in biological populations challenges fundamental assumptions at the heart of continuum models of population dynamics based on mean densities (local or global) only.  Individual-based models (IBM's) were introduced over the last decade in an attempt to overcome this limitation by following explicitly each individual in the population.  Although the IBM approach has been quite insightful,  the capability to follow each individual usually comes at the expense of analytical tractability, which limits the generality of the statements that can be made.  For the specific case of spatial structure in populations of sessile (and identical) organisms,  space--time point processes with \emph{local regulation} seem to cover the middle ground between analytical tractability and a higher degree of biological realism.  This approach has shown that simplified representations of fecundity, local dispersal and density--dependent mortality weighted by the local competitive environment are sufficient to generate spatial patterns that mimic field observations.  Continuum approximations of these stochastic processes try to distill their fundamental properties, but because they keep track of not only mean densities, but also higher order spatial correlations, they result in infinite hierarchies of moment equations. This leads to the problem of finding a `moment closure'; that is, an appropriate order of (lower order) truncation, together with a method of expressing the highest order density not explicitly modelled in the truncated hierarchy in terms of the lower order densities.   We use the principle of constrained maximum entropy to derive a closure relationship at second order using normalisation and the product densities of first and second orders as constraints, and apply it to one such hierarchy.   The resulting  `maxent'  closure is similar to the Kirkwood superposition approximation, or `power-3' closure,  but it is complemented with previously unknown correction terms that depend on integrals over the region for which third order correlations are irreducible. The region of irreducible triplet correlations is found as the domain that solves an integral equation associated with the normalisation constraint.  This also serves the purpose of a validation check, since a single, non--trivial domain can only be found if the assumptions of the closure are consistent with the predictions of the hierarchy.  Comparisons between simulations of the point process, alternative heuristic closures, and the maxent closure show significant improvements in the ability of the truncated hierarchy to predict equilibrium values for mildly aggregated spatial patterns.  However, the maxent closure performs comparatively poorly in segregated ones.  Although the closure is applied in the context of point processes, the method does not require fixed locations to be valid, and can in principle be applied to problems where the particles move, provided that their correlation functions are stationary in space and time.
\end{abstract}

\section{Introduction}
\label{sec:introduction}
One of the most widely used models in theoretical ecology is the logistic equation \cite{mur93,pearl20,ver38}
 \begin{eqnarray} 
  \label{eq:logistic}
  \frac{d}{dt} \ma(t)  &=& r\,\ma(t)\left( 1 - \frac{\ma(t)}{K} \right)  \\
  \nonumber \ma(0)  &=& n_0,
 \end{eqnarray}
which describes the dynamics of a population in terms of a single state variable $\ma(t)$, which can be interpreted as the total population size or as the global density.
  The rate of change of the density in the logistic model is determined by three drivers. The first two are present in  the net growth term $r=b-d$, where $b$ and $d$ are respectively the \emph{per capita} fecundity and intrinsic mortality rates.  The third one is the density-dependent mortality rate, which is assumed to be proportional to the density,  where the constant of proportionality $K$ is the `carrying capacity', i.e. the maximum number of individuals per unit area or volume that can be supported by some unspecified limiting resource. 
 %
 This model is built on the following set of assumptions  \cite{bol97,die00,law03}:
 \begin{enumerate}
   \item There are no facilitative interactions among conspecifics.    \label{enu:ass1}
   \item  Contributions to mortality due to competition are pairwise additive.     
             \label{enu:ass2}
  \item The limiting resource is uniformly distributed in space, and shared proportionally by all individuals. \label{enu:ass3}
  \item  There are no differences 
           among individuals in age, size or phenotype.
           \label{enu:ass4}
 \item The spatial locations of the individuals are uncorrelated. 
          \label{enu:ass5}
 \item Allocation to reproductive tissues is independent of the local resource availability.  \label{enu:ass6}         
\item Density--dependent mortality occurs at the same temporal scales than fecundity and intrinsic mortality. \label{enu:ass7}
\end{enumerate}
These assumptions are valid only for a rather restricted set of biological situations.  For instance, facilitative interactions are known to play a determinant role alongside competition in shaping community structure and dynamics \cite{brooker08}.  In plant communities,  non-succesional positive interactions can result from additional resources being  made available through synergies (e.g. hydraulic lift, microbial enhancement, mycorrhizal networks),  a reduction in the impact of climate extremes and predation \cite{gross08} or a combination of these.  The assumption of pairwise additivity in density-dependent mortality enjoys some degree of empirical support for plant populations \cite{weigelt07}, but it is still an unresolved issue \cite{damgaard07,dormann05}.  Forms of population structure driven by size (or age), phenotype or spatial pre-patterning in the abiotic substrate having an impact on fecundity, recruitment and survivorship are ubiquitously observed both in the field and experimental literature \cite{pur02,turnbul07} \cite{sil93}.  Seed dispersal and competitive interactions are known to occur over a characteristic range of spatial scales rather than being uniformly distributed as is commonly assumed in the logistic model 
\cite{dal99,gra04,sch06,sil93,silvertown00,sto00,con00}.\\

These limitations have motivated the search for alternatives to  the logistic equation that can address questions of broader biological interest, while simultaneously maintaining a reasonable degree of mathematical and computational tractability.  Achieving this goal depends heavily on the development of multiscale modeling approaches capable of linking patterns manifested at the larger, population--level scales, to their  drivers, which lie in biological processes occurring at the level of individuals;  typically taking place over spatial and temporal scales that differ substantially  from those at which the population--level regularities are detected \cite{bol97,bol03a,die00,dur99,law00,law03,moo01,lev94,sat00}. \\

Among all the possible paths suggested as one relaxes these assumptions (\ref{enu:ass1}--\ref{enu:ass7}),  understanding  the role of spatial structure, particularly that driven by biological processes alone,  has received a considerable amount of interest \cite{die00a,law03,bol97,bol99,bol00,scanlon07,dal99,borgogno09,iwa00}.  The  approaches that have been developed for the spatial problem have a number of commonalities. They usually consist of an individual-based model (IBM) \cite{dea05,gri99} which follows simplified representations of the life histories of each individual in the population.  These representations include the  biological processes believed to play a role in driving the population--level phenomena, and typically include a combination of
 fecundity, dispersal, mortality and in some cases, growth.  These are modeled in such a way that some  form of density--dependent regulation is present in at least  in at least one of them. Second, the density--dependent regulation is determined by the neighborhood configuration surrounding each focal individual, which leads to a \emph{local} regulation of the  process \cite{blath07,eth04,fou04}. Third, the dynamics of the macroscopic patterns is obtained from an average of a sufficiently large number of independent realisations of the individual--level model.  Insights about the emergence of various forms of population structure, in particular space, are gained as these broad scale patterns are allowed to vary with the characteristic scales that regulate the biological processes at the level of the individual organism \cite{bol97,law00,pac97,pas99,wil00}. \\
 
This approach, albeit insightful, restricts severely the statements that can be made about how the processes present across various scales interact to produce pattern, since typically there is an absence of a model condensing  the dynamics of pattern at the larger scale.  To circumvent this deficiency, several attempts to derive population--level models from the IBM have been introduced in the literature. In the context of spatial pattern in plant population dynamics \cite{bol97,law00,iwa00,scanlon07}, these models typically take the form of hierarchies of equations for relevant families of summary statistics where quantities in addition to the mean density capture spatial correlations among pairs, triplets etc, that quantify spatial pattern across a range of scales \cite{dal99,stoyan95}. These summary statistics are closely related to the central, factorial or  raw spatial moments of the underlying spatial stochastic process.  For pair configurations in plant population models, common choices are the spatial auto-covariance or the second order product density \cite{dal99,dal03,diggle83,stoyan95}.  A discussion of these various approaches in the development of continuum approximations to spatio-temporal stochastic processes in ecology can be found in a compilation edited by Dieckmann \emph{et al} \cite{die00a}. \\

The non-linearities due to the presence of density--dependence in spatially explicit IBM's inevitably result in infinite hierarchies of evolution equations for the summary statistics, where the dynamics of the correlations of order $k$ is tied to that of order $k+1$.  If one truncates the hierarchy at some order, the evolution equation at the order of the truncation will depend on the \emph{unknown} density of the next higher order. Analysis of these hierarchies can only proceed after truncation for some small order. This requires the solution of two problems. The first, is identifying an appropriate order of truncation $k$. The second is compensating for the resulting loss of information. The order of truncation in existing models is chosen on the basis of computational complexity, and rarely goes beyond two \cite{sin04,bol97,law00}.  For the second problem,  the density of order $k+1$ is replaced by a functional relationship of all the densities of order up to $k$, usually called a `moment closure'. This functional dependence of higher order quantities on lower order ones is constructed mainly on heuristic reasoning \cite{bol97,die00,mur04}. For instance,  when the order of truncation is two, assuming vanishing central moments of order three leads to the so-called `power--1' closure \cite{bol97}. The `power--2' closure arises from an analogy with the pair approximation used in discrete spatial models \cite{iwa00,die00}.  Assuming independence of the three pair correlations associated with each edge of a triplet for all spatial scales leads to the `power--3' or Kirkwood superposition approximation \cite{kir42,die00}. Although higher order closures do exist , they have restricted applicability due to the daunting computational problem that results at orders higher than three \cite{sin04}.\\

Despite some encouraging success that resulted in analytical solutions of the hierarchy at equilibrium for truncation at second order \cite{bol97,bol99,bol00}, and remarkably good fit of the numerical solution of the hierarchy with individual--based  simulations with so--called \emph{asymmetric} versions of previously used closures \cite{law03,mur04},  most predict poorly the equilibrium densities even for situations of mild spatial correlations.  In the cases where they succeed over a broader range of regimes of spatial correlations  (i.e. the asymmetric power--2), the closure depends on tuning  a set of weighting constants whose values can presently be found only by comparison with simulations of the stochastic process. A significant obstacle in the widespread adoption of these continuum approximations and their closures is that none of them is equipped with  a criterion for their domain of validity that does not depend on  comparisons with simulations of the individual--based model.  Nevertheless, many of these heuristic closures do provide a better approximation to the dynamics of a spatially structured population than the logistic equation, and illuminate a variety of mechanisms by which endogenously generated spatial pattern appears in plant populations. \\

Inspired by earlier results of Hillen \cite{hil04} and Singer \cite{sin04}, who  used the principle of constrained maximum entropy \cite{sha49,khi57} \cite{jay57}  to respectively derive closures for velocity jump processes \cite{othmer88} and the BBGKY hierarchy arising in the statistical mechanics of fluids \cite{kir42}, we develop a closure scheme based on constrained entropy maximisation for the moment hierarchy
developed by Law \& Dieckmann \cite{law00}, constrained to satisfy normalisation and the product densities up to order two.  In order to be able to relate the output of the entropy maximisation to the approximating dynamical system, we also reframe the hierarchy of Law \& Dieckmann \cite{law00} in terms of product densities rather than the spatial moments.  These two kinds of sets of summary statistics are very closely related, since the latter can be seen as estimators of the former.    The approach of Hillen \cite{hil04} consists of proving that the $L^2$--norm over the space of velocities of the transport equation of Othmer \emph{et al} \cite{othmer88}  behaves like an entropy, with the velocity moments acting as constraints. Singer \cite{sin04} treats the triplet product density as a probability density in order to construct an entropy from the point of view of information theory \cite{sha49,khi57,jay57}, using consistency of the marginals as constraints.  Our approach differs from these two other maximum entropy maximisation methods in a number of ways.  First, we use the information theoretical entropy functional for point processes \cite{fad65,dal04}, based on the negative of the expected log--likelihood, and includes all the orders that contribute spatial information, not just order three.  Second, the product densities which provide the constraints are incorporated into the entropy functional by means of an expansion that allows to express the likelihoods (or Janossy densities) in terms of product densities and vice versa \cite{dal88,dal03}, this allows us to establish a formal connection between the entropy functional and the moment hierarchy. Third, our closure is implicit, in the sense that the density of order three appears at both sides of the closing relationship, thus allowing irreducible correlations of third order to be explicitly included. Fourth, 
the method presented here complements the Kirkwood (or power--3) closure with previously unknown correction terms that depend on the area for which the three points in the triplet become independent.  These correction terms are important where the three particles in the triplet configuration are close to each other, but progressively vanish as these become separated, at which point the maximum entropy closure reduces to the classical Kirkwood superposition approximation.  These correction terms lead to substantial improvements in the prediction of the equilibrium density for mildly aggregated patterns. In addition, the closure comes equipped with a criterion of validity stemming from the normalisation constraint.  This validity check comes from an ancillary integral equation that returns the area of the domain at which the points become independent. This equation produces a single, non-trivial root when the correlations predicted by the moment hierarchy are consistent with the truncation assumptions, but fails to do so otherwise. 
The maximum entropy closure relationship we found is given by
\begin{eqnarray}
  \mc (\xi_{1},\xi_{2}) \label{eq:closurek=2intro}
 &=&
 \left[\mb(\xi_{1})-\,\ara \int_{A_0}\mc(\xi_{1},\xi_{2}')\,\,d\xi_{2}'\right] \\ 
 \nonumber &\times&
 \left[\mb(\xi_{2})- \ara \int_{A_0}\mc(\xi_{2},\xi_{2}-\xi_{1}')\,d\xi_{1}'\right]\\
 \nonumber &\times&
 \left[\mb(\xi_{2}-\xi_{1})-\,\ara \int_{A_0}\mc(\xi_{2}-\xi_{1}',\xi_{1}')\,d\xi_{1}'\right]\\
 \nonumber &\times& \frac{ J_{0}(A_0)}{\left[\ma-\,\ara \int_{A_0}
 \mb(\xi_{1}')\,d\xi_{1}'+\frac{\ara^{2}}{2}\int_{A_0\times A_0}
 \mc (\xi_{1}',\xi_{2}')\,d\xi_{1}'\, d\xi_{2}'\right]^{3}},
\end{eqnarray}
where $\ma,\mb,\mc$ are the first, second and third order product densities (the densities of the factorial moments of the underlying spatial point process), $\xi_1$, and $\xi_2$ are  vector distances respectively linking the pairs of particles $(x_{1},x_{2})$ and $(x_{1},x_{3})$ conforming a triplet configuration.  The set $A_0$ is a circular domain of area $|A_{0}|$ that establishes the spatial scale for which triplet correlations are irreducible, and $J_0(A)$ is the avoidance function (i.e. the probability of observing no points in $A$) of the spatial point process for the window $A$ \cite{dal88,dal03}.  This set is found as the domain of integration that solves the normalisation condition
\begin{equation}
  \int_{A_\epsilon}\mb(\xi_{1}')\,d\xi_{1}'-\frac{1}{3}\int_{A_\epsilon \times A_\epsilon}
  \mc(\xi_{1}',\xi_{2}')\,d\xi_1'\,d\xi_2'=|A_\epsilon|\, \ma^2-\frac{|A_\epsilon|^2}{3}\ma^3
  \label{eq:normalisation1}
\end{equation}
where $A_\epsilon$ is a circular domain of radius $\epsilon$ centered at the origin. The set $A_0$ is found by allowing the radius $\epsilon$ to take positive real values until  the equality in (\ref{eq:normalisation1}) holds. This closure is applied if the three points in the triplet lie inside $A_{0}$, and outside this region the classical Kirkwood closure applies.  If the area of normalisation $A_{0}$ is small, the largest correction is due to the $J_{0}$ term since the integral correction terms in the numerator and denominator tend to cancel each other, in which case the maxent closure is simply given by

\begin{equation}
\label{eq:closurek=2exp}
  \mc (\xi_{1},\xi_{2}) =
                     \frac{
                     \mb(\xi_{1}) \, \mb(\xi_{2})\,\mb(\xi_{2}-\xi_{1})
                          }{
                          \ma^{3}
                          } \exp(-\ma \ara)
 \end{equation}
where the exponential term corresponds to the avoidance function of a Poisson point process of mean density $\ma$ normalised with respect to the window $A_{0}$. \\


%
The paper is organized as follows.  Section \ref{sec:ppmodel} discusses the locally regulated space-time point process model originally developed by Law \& Dieckmann \cite{law00}, and includes a Gillespie-type simulation algorithm \cite{gil76,ren91},  together with known definitions and estimators for the product densities and some simulation results included for illustration purposes only.  Broader simulation results for this process can be found elsewhere \cite{law03,mur04,rag06}. Section \ref{sec:momentEquations} reframes the spatial moment equations of Law \emph{et al} \cite{law03} in terms of product densities. Section \ref{sec:entropyMax} discusses the moment closure for truncation at second order based on constrained entropy maximisation. Section \ref{sec:numerics} discusses the numerical implementation of the closure and compares its predictions against simulations of the point process for mildly aggregated patterns. Finally, section \ref{sec:conclusions} presents a critique of the maximum entropy method method, and suggests further areas of development.

\section{Spatio-temporal point process model}
\label{sec:ppmodel}
We consider a single population of identical individuals, each of which can occupy arbitrary locations on  a 2-dimensional continuous and bounded spatial arena $A$.  The state of the population for each fixed time $t$ is modeled as a realisation of a spatial point process, called the configuration or point pattern  
 \cite{dal03,stoyan95,diggle83},
  \begin{equation}
   \label{eq:configuration}
  \varphi_{t}({A})=\left\{x_1,\ldots,x_{N_t} \right\},
  \end{equation}
where  the $x_i$ are the spatial locations of all individuals found within $A$. Alternatively we have
\[
N_{t}(A)=\#\left\{x_1,\ldots,x_{N_t} \right\}
\] 
where $N_{t}(A)$ stands for the total population counts within $A$, and the cardinality operator $\#$ counts the number of elements in a set.  Note that  in (\ref{eq:configuration}) both the locations $x_i$ and the total counts $N_t$ are random variables. The dynamics of the population is modeled by introducing a time component, where the updating times are also random variables,  subject to \emph{local} regulation \cite{cre91,dal03}.   Two versions of this model have been   introduced independently by Bolker \& Pacala \cite{bol97} and Dieckmann \& Law \cite{die00}.  Both  share the key ingredients of non-uniform dispersal, and a density-dependent mortality term that depends on the configuration surrounding the focal individual which is the mechanism that introduces the local regulation. The configuration (\ref{eq:configuration}) evolves in time by sampling from two  exponential distributions of  waiting times that regulate the inter-event times between fecundity/dispersal and mortality events at the individual level, where the latter is determined from both intrinsic and density-dependent contributions.

\begin{table}[htdp]
\caption{Point process model parameters}
\centering
\label{table:pp_parameters}
\begin{tabular}{|lll|}
\hline 
Parameter                   & symbol             & units \\[3pt]
fecundity                    & $b$                & time$^{-1}$ \\
intrinsic mortality            & $\di$               &  time$^{-1}$ \\
density-dependent mortality  & $\dn$              & time$^{-1}$  indiv$^{-1}$ \\
non-spatial carrying capacity & $K$                &  individuals \\
dispersal scale              & $\sigma_\dis$ &  length \\
competition scale           & $\sigma_\mor$  & length \\
initial population size        & $\no$              & individuals \\
spatial arena                & $A$                & length$^2$\\ 
\hline
\end{tabular}
\end{table}%


  \subsection{Dispersal and fecundity}
   \label{ssec:birthRate}
     Per capita waiting times between births are assumed to be exponentially   
     distributed  with constant parameter $b$, the \emph{birth (or
     fecundity) rate}. 
     If a birth occurs, the newborn
     is displaced instantaneously from the location of its
     mother $x_i$ to a random new location $x_j$, sampled from the probability 
     density, $\dis(x_i-x_j;\sigma_\dis)$ 
     the \emph{dispersal kernel}, where $\sigma_\dis$ is a parameter that measures the characteristic 
     dispersal  length. The index $i$ of the mother is chosen uniformly from the list of indices $J_{A}=\{1,2,\ldots,N_{t}(A)\}$ in the configuration.
  \subsection{Mortality}
   \label{ssec:mortaRate}
     The probability that a given individual $i$ at location $x_{i}$ dies in the time interval
     $(t,t+dt)$ is  also assumed to be exponentially distributed with parameter $m(x_i)$, 
     the \emph{total per capita mortality rate},  given by 
     \begin{equation}
        m(x_{i}) = \di+\dn \sum_{j \neq i \,\in\, J_{A}} \mor(|x_{i}-x_{j}|;\sigma_\mor),
        \label{eq:densDepMortRate}
     \end{equation}
     where $\di$, is the \emph{intrinsic} mortality
     rate, and $\dn$ is the
     \emph{density--dependent} mortality rate. In order to allow comparisons with the predictions of the 
     logistic model (\ref{eq:logistic}) we defined it as $\dn =(b-\di)/K$, where $K$ is the non-spatial carrying 
     capacity 
     (the expected value at equilibrium under complete spatial randomness). This
     second `mortality clock'  is rescaled by a weighted average of  the local configuration around the focal 
     individual, so that mortality due to competition is more likely to occur in locally 
     dense regions than in comparatively sparse ones. The 
     contributions of neighbors to the mortality of $x_i$ are assumed to
     decay monotonically with distance.  This is modeled  by a normalized, radially 
     symmetric 
     weighting function $
     \mor(|\xi|\,;
     \sigma_\mor)$, the 
     \emph{mortality kernel}, that vanishes outside a finite interaction domain $D_\mor$
     , where $\sigma_\mor$  is a parameter associated with 
     the characteristic length scale of competitive interactions. 
     This function is interpreted as  an average effect that simplifies the details of the 
     physiology of mortality due to crowding.   The parameters of the model are 
     summarized in Table \ref{table:pp_parameters}.
  \subsection{Simulation algorithm}
   \label{ssec:simulation_algorithm}
     A sample path for the space-time point process with rates described in Sections  \ref{ssec:birthRate} and  \ref{ssec:mortaRate}   can be simulated by a variant of the 
     Gillespie algorithm \cite{gil76,ren91}. The spatial arena can be identified with the unit square $W=[0,1]\times[0,1]$ (after rescaling the parameters in the interaction kernels), with periodic boundary conditions.  The initial 
     population consists of $\no$ individuals, and $[0,T_{
     \mbox{max}}]$ is the time interval of interest.  
     \begin{enumerate}
        \item Generate  the configuration  at time $t=0,~  \varphi_{0}=\{ (x_
              {1},y_ {1}); \ldots ; (x_{\no},y_{\no}) \}$, from two independent sets of 
              $\no$ deviates from $U(0,1)$, $X_0=\{x_{1}, \ldots,x_{\no}\}$ and $Y_0=\{y_
              {1}, \ldots ,y_{\no} \}$.\\ 
        \item While the elapsed time $t$ is less than $T_{\mbox{max}}$ do:\\
          \begin{enumerate}
            \item Generate a  birth waiting time $T_{b}$
             from the exponential density with parameter $b\,N_t$, where $N_t$ is the 
             number of individuals that are alive at time $t$. 
                  \label{en:sim1} \\
            \item Generate the set of mortality waiting times $T_m =\{ \tau_1,
                  \dots,\tau_{N_t} \}$ from a set of exponential
                  densities, each with parameter $\mi=\di+\dn \sum_{j \neq i} \mor(|
                  x_{i}- x_{j}|)$,  for each of the $i=1,\ldots N_t$ individuals in the 
                  configuration at time $t$ \\
            \item The time until the next event is given by $\tau_{n}=\min\{T_{b} \cup 
                  T_m\}$. \\  
            \begin{enumerate}
               \item A birth occurs if $\tau_{n}=T_{b}$, in which case the location of the newborn individual $x_{b}$ is given by
                       \[x_{b} = x_p+\xi \]
                   where the index of the parent
                   $p$ is drawn uniformly from the set of indices $J_A$ and the 
                   displacement $\xi$ is drawn from the dispersal kernel $\dis(\xi)$. The configuration is then updated to include the newborn
                   \[                 
\varphi_{t+T_{b}} \rightarrow \varphi_{t} \cup\{x_{b}\} .   
\]          
              \item If $\tau_{n} \neq T_{b}$ then the next event is a death in which case the $i$-th 
                    individual in $T_{m}$ for  which $\tau_{i} = \tau_n$ is removed from the
                    configuration 
                     \[
                     \varphi_{t+\tau_{n}} \rightarrow \varphi_{t} \setminus \{x_{i}\}
                     \]
            \end{enumerate}
          \item Update the elapsed time  $t \rightarrow t+\tau_n$. 
                \label {en:sim2}
        \end{enumerate}
     \end{enumerate} 
  \subsection{Summary statistics}
   \label{ssec:summary}
\begin{figure*} 
         \centering
         \includegraphics[width=\textwidth]{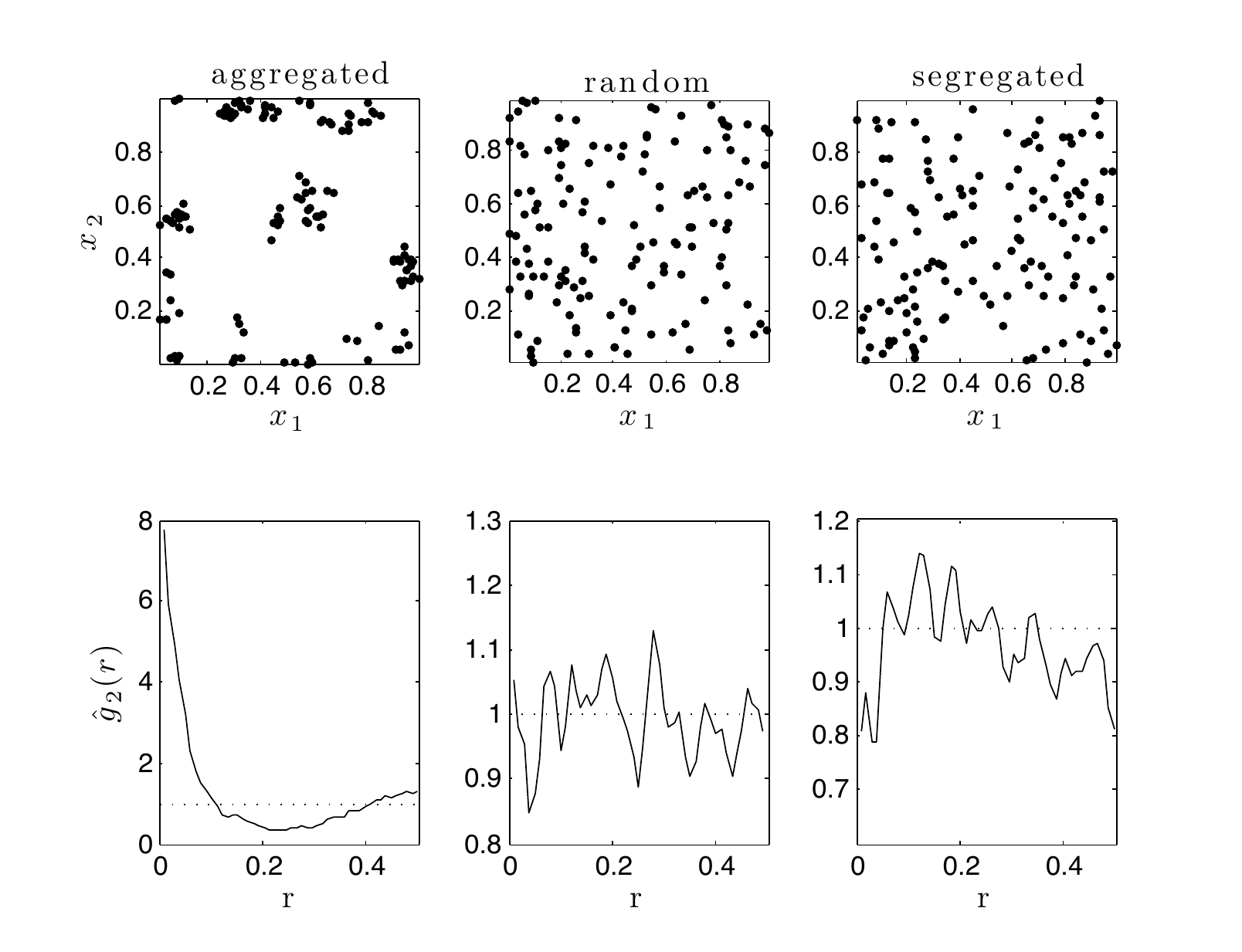}
         \caption{The three upper panels show different types of 
         point patterns sharing the same number of points $N(A) = 136$, where the window $A$ is the unit square.
         The left panel shows aggregation, the center panel corresponds to complete spatial randomness  and the 
         the right panel displays  a segregated  pattern. In the 
         aggregated pattern we see the tendency of points to occur near each other.  By 
         contrast 
         in the regular pattern points tend to avoid each other at short spatial 
         scales.  The lower three panels show estimates of the pair correlation function $\hat{g}_2(r)$ for each of the three point patterns at the top.  The lower left panel indicates aggregation at short scales but segregation at intermediate ones.  In the lower center panel the pair correlation function oscillates rapidly around one, which signals randomness, and the lower right panel indicates a tendency to segregation at short scales.
         \label{fig:pointPatterns}
         }
       \end{figure*}

  The specific configurations resulting from simulations of the algorithm in Section \ref
     {ssec:simulation_algorithm} are of limited interest.  The fundamental question is understanding how  spatial correlations develop from an unstructured initial condition, and how the equilibrium density departs from the logistic behavior when considering an ensemble of simulations for various 
     combinations of the spatial scales of competition and dispersal
     \cite{bol97,law00,die00}. This requires a set of summary statistics  capable of distinguishing various forms of 
     spatial structure for the same population size (see Figure 
      \ref{fig:pointPatterns}).  
      A useful set for this task are the product densities (or 
     densities of the factorial moments), i.e
       the densities of the expected 
     configurations involving one, two or more \emph{distinct} points after removing self-configurations
     \cite{stoyan95,dal03,diggle83}.  For spatially stationary point processes, 
     these are functions of the inter--point distances between 
     the points comprising an expected
     configuration of a certain order $k$.   The product densities are defined in terms of  
     the population count  
     $N_{t}(B)$ observed through some window $B$ at time $t$  defined as 
     \cite{stoyan95,cre91,dal03} 
  \begin{equation}
     N_{t}(B)=\sum_{x_{i} \in \varphi_{t}}I_{B}\,(x_{i}),
     \label{eq:count}
  \end{equation}
where $I_{B}(x)$ is the indicator function of the set $B$ defined by
  \begin{eqnarray}
      I_{B}(x)=\left\{\begin{array}{l}1~~ \mbox{if}~ x \in B,
      \\  0~~ \mbox{otherwise.}\end{array}\right.
      \label{eq:indicatorFunction}
    \end{eqnarray}  
     The coarsest is the mean density (or intensity) which measures 
     the expected  number of individuals per unit area at each time, defined as
      \begin{equation}
        \ma(x\,,t)=\lim_{ \epsilon \downarrow 0 }\frac{\E \{ N_{t}(\,S_\epsilon (x)\,)\}}{|S_\epsilon(x)|}        \label{eq:firstMomDef}
      \end{equation} 
     where $S_\epsilon (x)$ is the open ball of radius $\epsilon$ centered around $x$, and  $|A|$ is the area of the window $A$.  Since the mortality and fecundity rates do not depend specific locations but on relative distances, and both  the dispersal or competition kernels are symmetric by definition, the spatial point process is spatially stationary and isotropic, in which case the mean density is constant for each fixed time
     \[
     \ma(x\, ,t)=\ma(t).
     \]
     A na\"ive estimator for the mean density from a single realisation is \cite{stoyan95,diggle83} 
      \begin{equation}
       \hat{m}_1(t)=\frac{N_{t}(A)}{|A|}
       \label{eq:firstEstimatora}
      \end{equation}
     where $N_{t}(A)$ is as in (\ref{eq:count}).  
     If an ensemble of $\Omega$ independent replicates of the process is available, this 
     estimate can be improved by averaging over the ensemble 
     \begin{equation}
       \bar{m}_1(t)=\frac{\langle N_{t}(A) \rangle_\Omega }{|A|}.
       \label{eq:firstEstimator}
      \end{equation}
      For a Poisson process, the mean density 
     (\ref{eq:firstMomDef}) is a sufficient statistic for the process.  More general cases 
     require keeping track of spatial correlations. Higher order quantities 
     are required to distinguish between aggregated (or clustered), random and segregated (or over--dispersed) point patterns
     with the same mean density (see Figure \ref{fig:pointPatterns}).  For this purpose we 
     need, at the very least, information about two-point correlations.  These are measured by the pair correlation function, 
     defined as the ratio
      \begin{equation}
       g_{2}(\xi\,;t) = \frac{ \mb(\xi,t) }{ \ma^{2}(t) }
       \label{eq:pairCorrFunDef}
      \end{equation}
     which requires knowledge of the density of the expected number of pairs at spatial 
     lag $\xi$, measured by the second order product density  $\mb(\xi,t)$. 
     \begin{equation}
         \mb(\xi\,;t)=\lim_{\epsilon \downarrow
          0}\frac{\E\left\{N_{t}(S_\epsilon(\mathbf{0})\,)\left[\,N_{t}(S_\epsilon(\mathbf{0}+\xi)\,)-\delta_{\mathbf{0}}(S_\epsilon(\mathbf{0}+\xi)) \, \right]\right\}
          }{|S_\epsilon(\mathbf{0})|\, |S_\epsilon(\mathbf{0}+\xi)|}
          \label{eq:secondMomDef}
     \end{equation}
     where $S_\epsilon(\mathbf{0})$ and $S_\epsilon(\mathbf{0}+\xi)$ are small windows of observation respectively centered at the origin, and at distance $\xi$
     from the origin. 
     The Dirac measure in the second factor in the numerator removes the count at zero lag   
     from the second window in order to avoid self-configurations. In general, the definition 
     (\ref{eq:secondMomDef}) centers the count for each specific location $x$, but given that in our case the process is stationary and isotropic by construction, it can be translated to the origin without loss of generality, in which case
     $\mb$ depends  only on the spatial lag $\xi$. \\
        
      In the case of a spatially random configuration (a Poisson point process), the counts on non-overlapping windows are independent of each other and thus the second order density is simply the square of the mean density.  Correlations of configurations involving $k$ points are simply the $k$-th  powers of the mean density \cite{diggle83,stoyan95}.
       The pair correlation function (\ref
     {eq:pairCorrFunDef})
     is the lowest order product density that allows detection of departures from complete spatial randomness. Thus,  values of the pair correlation function greater than one 
     for some lag $\xi$ indicate aggregation at that scale, whereas values below one
    signal segregation.  Estimation of the pair correlation function 
    requires an estimator of the squared density  \cite{stoyan95}
    \[
    \bar{m}_1^{2}(t) =\frac{\langle \,N_{t}(A)\,[N_{t}(A)-1]\,\rangle_{\Omega}}{|A|^{2}},
    \]  
    together with a kernel density estimator for the second order product density \cite{scott92,stoyan94},
     \begin{equation}
       \widehat{m}^{\,(h)}_{2}(r,t) =\frac{1}{2 \pi r} \sum_{i}\sum_{j \neq i} \frac{ k_{h}(r-\|x_
       {i}-x_{j}\|) }
       {\,\left|A_{x_{i}} \cap A_{x_{j}}\right|} 
      \label{eq:secondEstimator}
     \end{equation} 
    where $r$ is the spatial lag, $h$ is the bandwidth of the kernel density estimate 
    $k_h$, the points $x_{i}$ belong to a configuration $\varphi_{t}(A)$ sampled 
    at time 
    $t$, and $\|x_i-x_j\|$ is the Euclidean distance between the points $x_i$ and $x_j
    $. The   
    denominator is an edge corrector that rescales the count in the numerator by the area of the intersection of the window of observation $A_{x_i}$ shifted so that its centered around the point $x_i$, with the window $A_{x_j}$ shifted around $x_j$ \cite{cre91,dal99,stoyan95}
\[ 
A_{x_{i}}=\{x+x_{i} : x \in A\}.
\]  
  If an ensemble of independent realisations is available, the single realisation  estimator (\ref{eq:secondEstimator}) can be improved by means of an ensemble average
    \[
    \bar{m}^{\,(h)}_{2}(r,t) =  \left< \widehat{m}^{\,(h)}_{2}(r,t) \right>_\Omega.
    \] 
    As 
    before, the angle brackets $\left< \right>_\Omega$ represent an average 
   of  the 
    estimates across a number of independent sample paths $\Omega$.  
     For the smoothing kernel $k_{h}$ a common choice is the 
    Epanechnikov kernel
     \[
       k_{h}(s)=\frac{3}{4 h}\left(1-\frac{s^{2}}{h^{2}}\right)I_{(-h,h)}(s),
     \]
    where $I$ is the indicator function (\ref{eq:indicatorFunction}).  Although empirical 
    methods for selection of the bandwidth $h$ are widely used, for instance the rule  
    \cite{stoyan94}
    \[
    h=c/\sqrt{\hat{m}_1(t) }, \,c \in (0.1,0.2), \] 
    data-driven methods for 
    optimal choices of $h$ based on cross-validation have been recently introduced \cite
    {guan07,guan07a}.  
In general, the product density of order $k$ is defined as \cite{bla95}
	\begin{eqnarray}
		&&m_{k}(x_1,\ldots,x_k,t) = \lim_{\epsilon \downarrow 0 }	
		\,\E
		\left\{
		             \prod_{j=1}^k \frac{
		                                       \left[
		                                        N_{t}(S_\epsilon (x_j))
		                                        -\sum_{i=1}^{j-1}\delta_{x_i}(S_\epsilon (x_j))
		                                        \right]
		                                        }{
		                                        | S_\epsilon (x_j) |
		} \right\}, 
		\label{eq:genDefProdDens}
	\end{eqnarray}
	where $\sum_{i=1}^{j-1}\delta_{x_i}(S_\epsilon (x_j))$ removes self
	$j$-tuples for $j>i$. In the case of spatial stationarity and isotropy, the specific 
	locations 
	$x_1, \ldots, x_k$ can be replaced by the relative distances $\xi_1,\ldots,\xi_{k-1}$, 
     \[
      \nonumber m_{k}(\xi_1,\ldots,\xi_{k-1},t),
     \]
    and the $k$-th correlation function becomes,
    \begin{equation}
     \nonumber g_{k}(\xi_1,\ldots,\xi_{k-1};t)=\frac{ m_{k}(\xi_1,\ldots,\xi_{k-1},t)}{m_1
     ^{\,k}(t)}
     \label{eq:kcorrfuncDef}
    \end{equation}
   which is interpreted in a similar way to the pair correlation function, but 
   considering $k$-plets instead of pairs.  \\
\subsection{Point process simulation results}\label{ssec:simulation}   
\begin{figure*} 
         \centering
         \includegraphics[width=\textwidth]{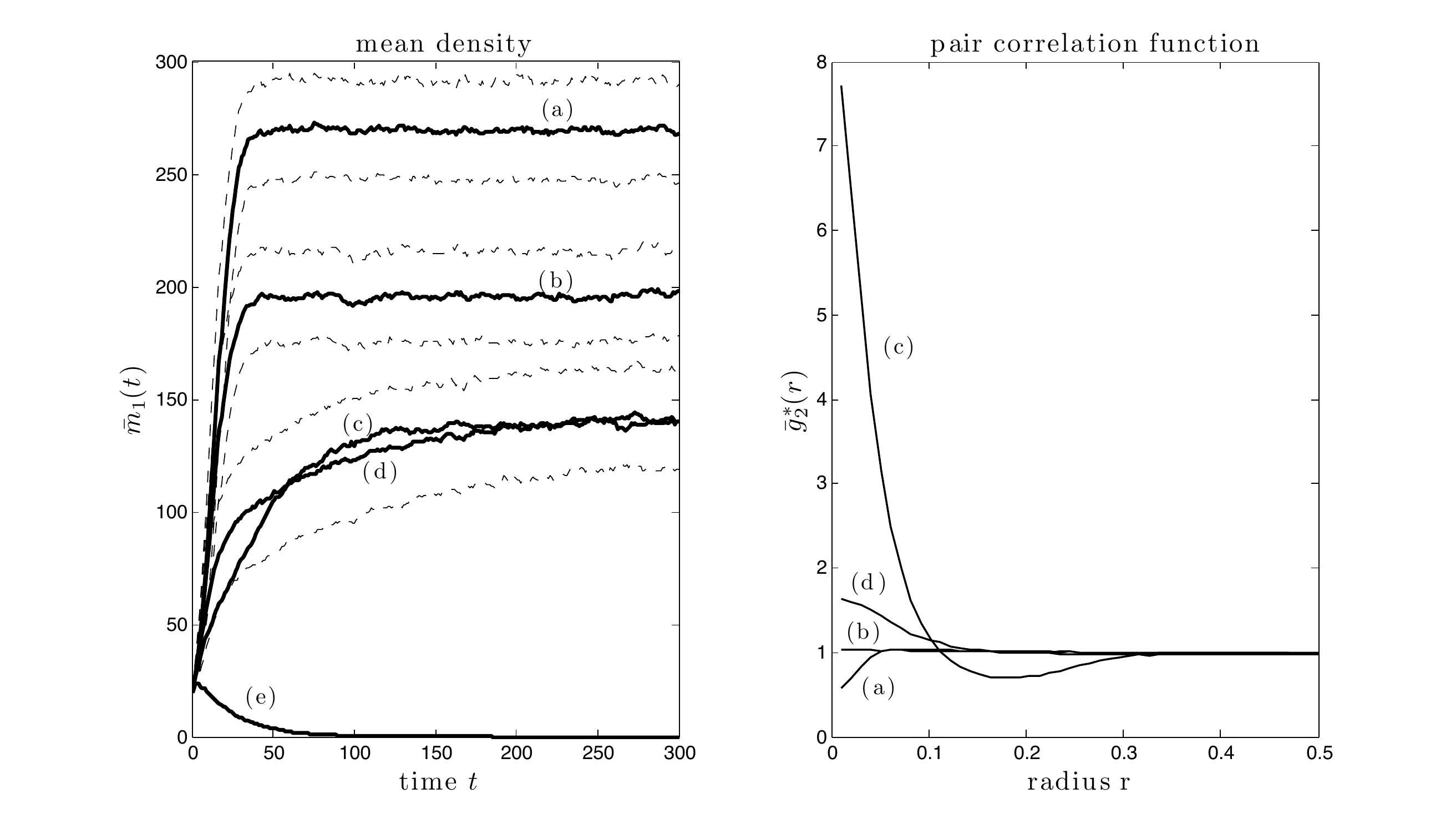}
         \caption{The left panel shows estimates for the mean density $\bar{m}_1(t)$ from an ensemble of $\Omega=300$ realisations, for various characteristic spatial scales of dispersal and density-dependent mortality. The dotted lines are the envelopes for one standard deviation. The right panel shows the corresponding estimates for the pair correlation function $\bar{g}_{2}^{\ast}(r)$ at the end of the simulation. The other parameters, $b=0.4,\,d=0.2,\,K=200$, are fixed for all cases. The spatial arena is the unit square with periodic boundaries. 
         \label{fig:firstandsecondsim}
         }
       \end{figure*}

 For the convenience of the reader, simulation results for the point process are shown in Figure \ref{fig:firstandsecondsim}, with the same parameter values as in Law \emph{et al} \cite{law03}, but obtained from code developed independently.  The spatial arena is the unit square, and the kernels are both radially symmetric Gaussians, but the mortality kernel is truncated (and renormalized) at $3\, \sigma_{\mor}$. The left panel shows estimates of the mean density versus time for various values of the characteristic spatial scales of dispersal and mortality.  The right panel shows the pair correlation function at the end of the simulation for each of the four spatial regimes for which the population persists.  Both quantities were estimated from an ensemble of 300 independent sample paths. \\

Case (b) in both panels corresponds to dispersal and mortality kernels with large characteristic spatial scales ($\sigma_{\dis}=0.12,\, \sigma_{\mor}=0.12$).  In this situation there is enough mixing to destroy spatial correlations ---confirmed by the almost constant pair correlation function--- and the mean density equilibrates at a value that is very close to the non-spatial carrying capacity ($K=200$).  Case (a) shows results for a segregated  (or regular) spatial pattern that arises from very local competitive interactions, but long range scales of dispersal ($\sigma_{\dis}=0.12,\, \sigma_{\mor}=0.02$).  In this situation local densities experienced by the focal individual are lower than the random case (the pair correlation function is below one), which results in equlibrium densities that equilibrate at higher values than the non-- spatial carrying capacity.  This results from the ability of newborns to escape locally crowded regions via the long range dispersal kernel.  Case (c) is associated to a segregated pattern of clusters, which is the converse situation of the segregated pattern with very localized dispersal, and mild competition distributed over a longer range ($\sigma_{\dis}=0.02,\, \sigma_{\mor}=0.12$).  The oscillations of the pair correlation function indicate two scales of pattern.  There is short scale aggregation, but the clusters themselves form a segregated pattern with respect to each other, so the local crowding due to clustering that should lead to high density-dependent mortality is compensated by the overdispersion.  Overall, the local competitive neighborhood experienced by an individual in this situation is more crowded than in a random distribution of points, which results in a mean density that equilibrates at lower values than the non-spatial carrying capacity.  Case (d) corresponds to a mildly aggregated pattern ($\sigma_{\dis}=0.04,\, \sigma_{\mor}=0.04$), where there is a single scale of aggregation.  Even for small departures from complete spatial randomness such as this one, the effect of the spatial pattern in the dynamics of the mean density is substantial, since we see a reduction of about $30\%$ in the equilibrium density in this case with respect to that of complete spatial randomness.  Finally, case (e) indicates an extreme case of aggregation, with very intense, local mortality and dispersal ($\sigma_{\dis}=0.02,\, \sigma_{\mor}=0.02$), where the population goes to extinction (exponentially) after a short growth transient.    
%
\section{Moment equations and the closure problem}
\label{sec:momentEquations}
The central problem associated with  the space-time point process described earlier in Section \ref{ssec:simulation_algorithm} is to obtain a closed form expression for the finite dimensional distributions, 
 \begin{equation}
  \label{eq:fidik}
  \pr_k\left\{ A_1,\ldots, A_k,n_1,\ldots,n_k;t\right\},
 \end{equation}
that determine the probability of observing $n_1$ points in the window $A_1$, $n_2$ points in the window $A_2$, and so forth up to the $n_k$ points in $A_k$ at time $t$, from the definition of the space-time point process discussed in the previous section.  Unfortunately, this seems to be remarkably difficult, due to the presence of the non-linearity in the mortality rate in (\ref{eq:densDepMortRate}), and the localized nature of dispersal \cite{eth04}.   However, the question of ecological interest is understanding the modifications that should be introduced to the logistic equation (\ref{eq:logistic}) in order to account for the effects of spatial correlations in the dynamics of the mean density. This can be accomplished by deriving evolution equations for the product densities (which are the densities of the factorial moments of (\ref{eq:fidik})) from the transition rates of the point process discussed in the previous Section.  Following a Master equation approach similar to that used by Bolker \& Pacala \cite{bol97} and Dieckmann \& Law \cite{die00}, we derive the following hierarchy of product density equations (see Appendix A). The first member in this hierarchy corresponds to the modified or `spatial' logistic equation \cite{mur92}, 
 \begin{eqnarray}
  \frac{d}{dt}\ma(t)=r\, \ma(t)- \dn \int_{\R^2}
   \mor(\xi_1)\,\mb(\xi_1',t)\,d\xi_1'.
  \label{eq:firstProdDensSpatr1}
 \end{eqnarray}
where $r= b-\di$, $\dn =(b-\di)/K_s$ and $\mor(\xi_1)$ is the mortality kernel in (\ref{eq:densDepMortRate}). $K_s$ is the $\emph{spatial}$ carrying capacity, or the number of individuals per unit area that can be supported under random mixing 
\[
K_s = \frac{K}{|A|}.
\]
Equation (\ref{eq:firstProdDensSpatr1}) shows that the required modification  of the logistic equation consists of substituting the quadratic term with an average of the second order product density $\mb(\xi_1,t)$ weighted by the mortality kernel $\mor(\xi_1)$.  This term computes the effective number of neighbors $n_{\mbox{eff}}$ that contribute to density--dependent mortality,
\[
n_{\mbox{eff}}\,(t)=  \int_{\R^2}
   \mor(\xi_1)\,\mb(\xi_1',t)\,d\xi_1'.
\]
 Thus, the effect of mortality on the evolution of the mean density is tied to a  weighted average of the mortality kernel with the two-point spatial correlations in the process.   Equation (\ref{eq:firstProdDensSpatr1}) reduces to the logistic equation for the Poisson point process, in which case $\mb(\xi_1)=\ma^2$.  In aggregated spatial patterns, $\mb$ exceeds $\ma^2$ for some domain.  If mortality is modeled by a kernel that penalizes close proximity over the same range of scales where aggregation is detected, then the effect of mortality due to competition is stronger in this case than that of the logistic equation, in which case the density equilibrates below $K_s$ (Figure \ref{fig:firstandsecondsim}, cases (c),(d) and (e) ). The opposite situation occurs in segregated patterns, where $\mb$ is less than $\ma^2$ at the scales  where the mortality kernel penalizes aggregation. As a result, the effect of competition on mortality is milder than in a random spatial pattern, in which case the mean density equilibrates at values greater than $K_s$ (Figure \ref{fig:firstandsecondsim}, case (a)).  Equation (\ref{eq:firstProdDensSpatr1}) depends on the unknown second order density $\mb$.  A similar procedure to that used in the derivation of (\ref{eq:firstProdDensSpatr1}) one obtains the evolution equation for this quantity
 \begin{eqnarray}
  \nonumber \frac{1}{2}\,\frac{d}{d
   t}\mb(\xi_{1},t)&=& b \int_{\R^2}
   \dis(\xi_{2})\,\mb(\xi_{1}-\xi_{2},t)\,d\xi_{2}+b\,\dis(\xi_{1})\,\ma(t)-\di\,
   \mb(\xi_{1},t) \\  &-&
   \dn \mor(\xi_{1})\,\mb(\xi_{1},t)-\dn\int_{\R^2}
   \mor(\xi_{2})\,\mc(\xi_{1},\xi_{2},t)\,d\xi_{2}.
   \label{eq:secondProdDensSpat2}
 \end{eqnarray}
 \begin{figure} 
  \centering
  \includegraphics[width=0.9\textwidth]{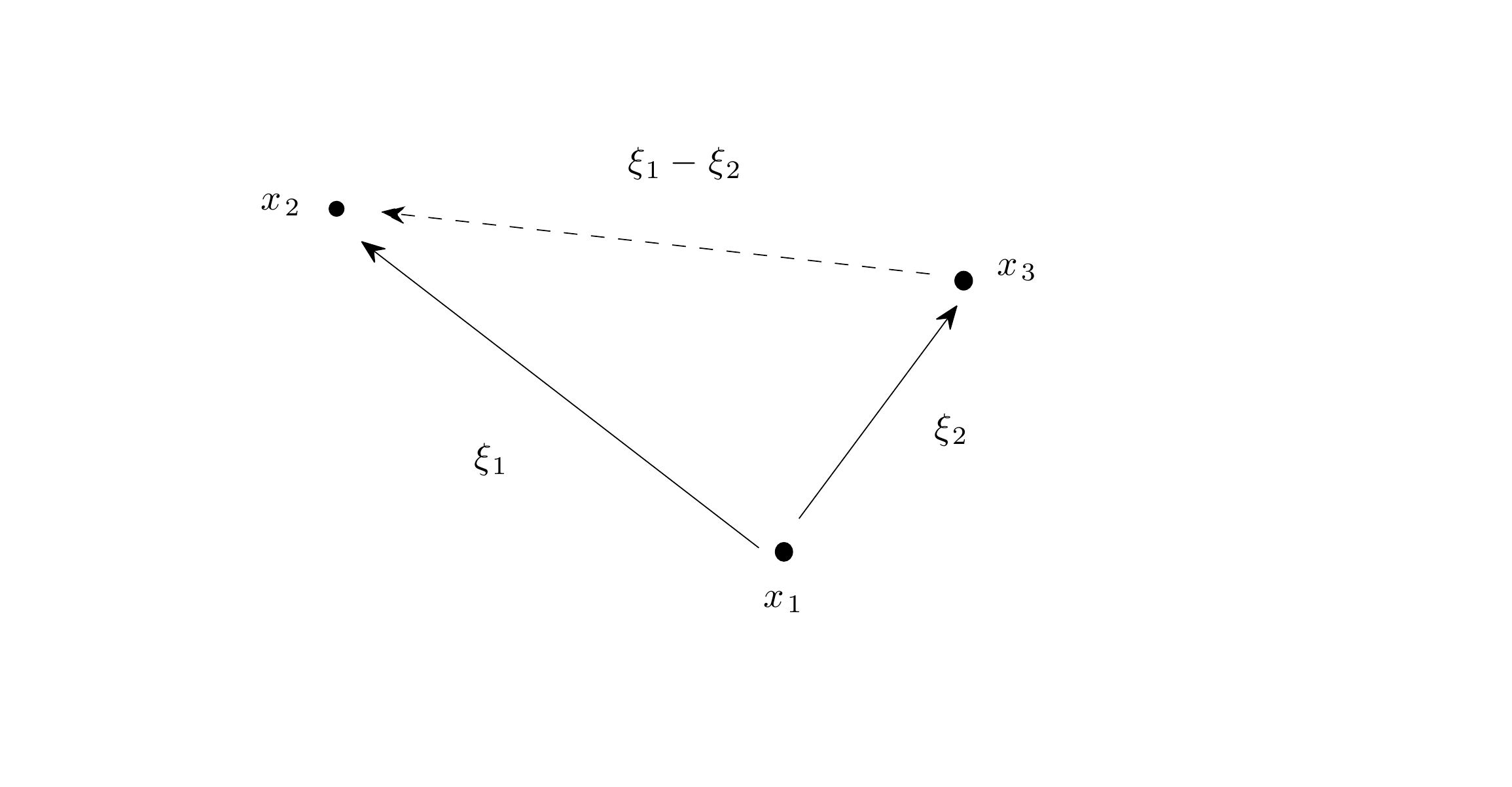}
  \caption{Schematic representation of a spatially stationary
                triplet configuration. The pair densities are evaluated at each 
                inter-event (vectorial) distances $\xi_{1}$, $\xi_{2}$ and $\xi_1-\xi_2$
                \label{fig:triplet}
                }
 \end{figure}
Here the role of dispersal and competition kernels as the main pattern drivers can be clearly discerned \cite{bol97,bol00,die00,law03}.  The first two terms in (\ref{eq:secondProdDensSpat2}), related to fecundity and dispersal, are
 \[
  b \int_{\R^2}
  \dis(\xi_{2})\,\mb(\xi_{1}-\xi_{2};t)\,d\xi_{2}+b\,\dis(\xi_{1})\,\ma(t).
 \]
Both are nonnegative by definition for all values of $\xi_1$ and $t$. The rate of change of $\mb$ increases due to their effect, and thus they drive aggregation at the scales controlled by the characteristic spatial scale of the dispersal kernel.   The convolution  measures the creation of pairs along $\xi_1$ due to dispersal of the third member of the triplet along the $\xi_1-\xi_2$ edge (Figure \ref{fig:triplet}).  The second term measures the creation of pairs along the $\xi_1$ edge due to the dispersal events generated the individual at the origin of $\xi_1$. The remaining terms due to mortality are,
 \[
  -\di\, \mb(\xi_{1},t) -
  \dn \mor(\xi_{1})\,\mb(\xi_{1},t)-\dn \int_{\R^2}
  \mor(\xi_{2})\,\mc(\xi_{1},\xi_{2};t)\,d\xi_{2}.
 \]
All the three terms are negative, and thus contribute to the destruction of pairs along the $\xi_1$ edge, leading to segregated patterns.  The first term measures intrinsic mortality of both members of the pair and the remaining ones are related to density--dependent mortality.  The second, measures mortality of the pairs due to competition at the scales controlled by the mortality kernel.  The last term measures the destruction of the pair along the $\xi_1$ edge due to the effect of competition with the additional member of the triplet located along the $\xi_2$ edge. \\

These terms for both dispersal and mortality are initially calculated by fixing the count at the origin of $\xi_{1}$ and let the count at the end of $\xi_{1}$  vary according to the fecundity, dispersal and mortality terms.  Symmetry considerations require consideration of the reverse situation, where the count at the end of $\xi_1$ is fixed, and the origin is allowed to vary.  Since these are symmetric, these additional terms lead  to the factor of $1/2$ on the left hand side of  the equation for the second order product density.
\section{Moment closure by Shannon entropy maximisation}\label{sec:entropyMax}
   The product density equations (\ref{eq:firstProdDensSpatr1})  and
   (\ref{eq:secondProdDensSpat2}) 
   cannot  be 
   solved in that form because the 
   evolution equation for the second order density has a  mortality term that 
   depends on a weighted average of the third order one.   Although it is 
   possible to derive an additional 
   evolution equation for this  quantity, it will involve an unknown fourth order 
   term, leading to a system that is not closed.  In general, the evolution equation for 
   the density of order $k$ will depend on the density of 
   order $k+1$. This gives rise to two problems,  known together as  `a moment closure' \cite{bol97,law00}. 
   The 
   first is choosing an appropriate order of truncation $k$, and the second is finding an 
   expression for the product  density of order $k+1$ in terms of the densities of orders 
   up to $k$ 
   (or  $k+1$ in the case of an implicit closure).  \\
   
   Ideally, the order of the 
   truncation should be based on an understanding of the convergence 
   properties of the hierarchy in order to establish error bounds.  
   In practice, the 
   order of the truncation is determined by the
   computational cost of the numerical solution, which is determined by the size of the arrays that 
   can be stored and operated on efficiently. Explicit representation of third order terms already  
   requires 
   least 
   $3.2$ Gb of memory using double precision and a relatively coarse discretisation of 
   ~100 grid points per dimension.  This situation pretty much constrains to three the highest
   order density that can be represented explicitly.\\
   
   From an applied perspective, the 
   first and second order terms are of greatest interest, since these respectively encode the dynamics 
   of the average density and the spatial covariance. The latter can be interpreted biologically as the 
   average environment experienced by an individual as a function of spatial
   scale \cite{law00,law03}. The shape of the second order correlation function can be used to 
   distinguish between aggregated, random and segregated spatial patterns sharing the same average density  
   (see Section \ref{ssec:summary}). \\
       
   Closure problems are pervasive in the statistical mechanics of fluids where thermodynamic 
   quantities are derived from the statistical properties of the particle distributions \cite{sin04}\cite
   {salpeter58,grouba04,sese05,meeron57,kir42}.  Here our intent is somewhat  
   similar in the sense that a detailed individual-based model is  
   used to inform a mean-field model that does not neglect the role of spatial fluctuations in density 
   due to endogenously generated spatial structure
   structure \cite{bol97,bol99,law03}. 
Within  spatial ecology, moment closures have been proposed with varying degrees 
   of success, using a  suite of methods, among which we have:
   \begin{itemize} 
    \item \emph{Heuristic reasoning}, where consistency arguments are used to construct 
   closing relationships \cite{law03,die00,mur04,bol97}.
    \item \emph
   {Distributional properties}, where closures are based on assuming a functional form for the  distribution of the process \cite{kri05}.
   \item \emph{Variational} methods, where it is assumed that the unknown distribution 
   optimizes some meaningful functional, usually an entropy--like object \cite{hil04,sin04}
   \end{itemize} 
   In order to make the paper reasonably self-contained, we shall briefly review closures based on heuristic reasoning, which have dominated work in this problem. Additional information can be found in a recent review by Murrell \emph{et al} \cite{mur04}.
\subsection{Heuristic methods of moment closure} \label{ssec:heuristic}   
Heuristic closures are usually based on self--consistency arguments.  For instance, they should be strictly positive and invariant under permutations of the arguments \cite{diggle83,cre91,dal03}. Also, if correlations are assumed to decay monotonically with distance, then there is a distance $d$ beyond which the particles become uncorrelated and thus higher order densities become simple powers of the mean density. 
Although a large number of functional forms can be chosen in order to satisfy these minimum requirements, the simplest ones usually involve additive combinations of various powers of the second and first moments.   For instance, if one further assumes that  central third moments vanish,  the resulting expansion in terms of product densities, leads to the \emph{power--1} closure, dubbed that way because the highest occurring power of the second order density is one \cite{bol97,bol99,bol00,die00}, 
  \begin{figure} 
   \includegraphics[width=\textwidth]{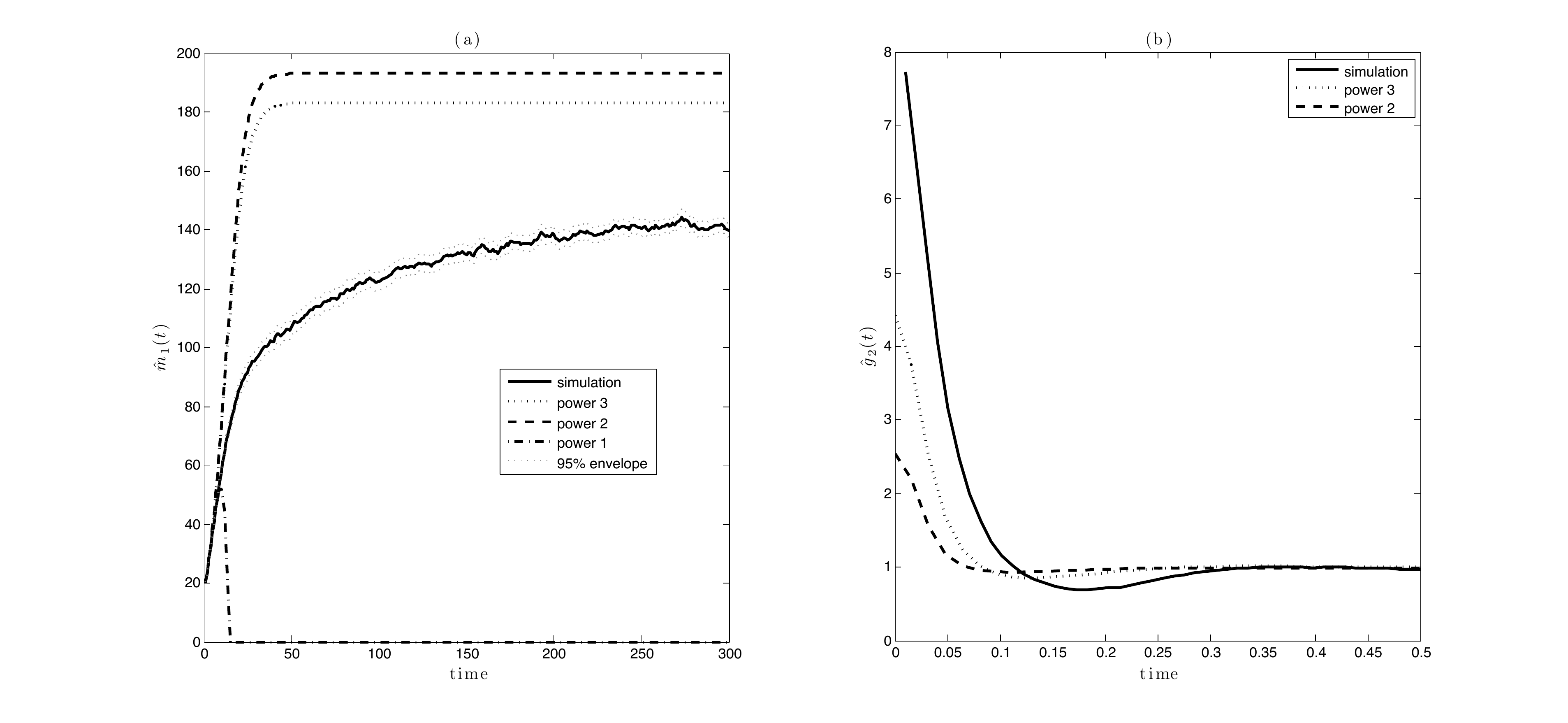}
   \caption{
     Closure comparison. Panel (a) shows the mean density ${\hat{m}}
    _1(t)$ of the point process versus time averaged over 300 sample paths (blue) up   
    to a simulation of 300 time units.  The continuous black line shows the predicted 
    mean density from the moment equations with the power--3 or Kirkwood closure, the 
    dashed black line corresponds to the power 2 closure.  The dash-dot line 
    corresponds two the power 1 closure.  Panel (b) shows the pair correlation function 
    at time $t=300$ (blue), indicating aggregation at short scales, but segregation at 
    intermediate ones.  The black line corresponds to the pair correlation function 
    predicted by the solution of the moment hierarchy with the power--3 closure, and the 
    dashed line corresponds to the power 2.
    \label{fig:compOldClosuresLat}
    }
 \end{figure}   

   \begin{equation}
   \label{eq:power1}
   \mc(\xi_{1},\xi_{2}) = \ma \, \mb(\xi_{1}) + \ma \,\mb(\xi_{2}) +\ma \, \mb(\xi_{1}-\xi_{2})-2\,\ma^3.  
   \end{equation}
This closure has the attractive property of preserving the linearity of the moment hierarchy, which allows the derivation of analytical results at equlibrium \cite{bol97,bol99}.  It is quite successful at low densities ($\ma^{\ast}\sim 20$) and 1--dimensional systems. However, at intermediate to high densities ($\ma\sim >100$)  aggregated patterns, this closure predicts extinction in situations where the point process persists (see dash-dot line in panel (a) in Figure \ref{fig:compOldClosuresLat}), even for mild correlation regimes.  It is nonetheless a  useful benchmark result.  \\

The \emph{power--2}  closure is obtained as a continuous space analogue to the pair approximation used in  discrete spatial systems \cite{sat00},
\begin{equation}
   \label{eq:power2}
   \mc(\xi_{1},\xi_{2}) = \frac{\mb(\xi_{1}) \,\mb (\xi_{2})
                                            }{ \ma}
                    	        +  \frac{\mb(\xi_{1}) \,\mb(\xi_{1}-\xi_{2})
                               		 }{ \ma} 
                   		+  \frac{\mb(\xi_{1}-\xi_{2})\,\mb(\xi_{2})
                               		}{\ma} 
                     		- 2 \,\ma^3;                                                         
\end{equation}
this closure does predict a persisting population.  However, it underestimates quite strongly the second order density, which leads to overshooting the mean density (see panel (b) in Figure \ref{fig:compOldClosuresLat}, dashed black line). It is non-linear and thus solutions have to be obtained numerically.  There are \emph{asymmetric} versions of this closure that consist of rescaling each additive term in (\ref{eq:power2}) with a set of weighting constants \cite{law03,mur04}.  Law \emph{et al} \cite{law00} showed that a particular combination of weighting constants provides a very good fit to simulations.  However, this result is difficult to generalize as there is no theory informing how these constants are chosen, since they depend on the details of the model \cite{mur04}, and can only be found by comparisons with simulations of the IBM. \\

Finally, the \emph{power--3} or Kirkwood closure (\ref{eq:power3}) has a distinguished tradition in the statistical mechanics of fluids \cite{kir42,kir42}. Recently, Singer \cite{sin04} showed that this closure can be obtained in the hydrodynamic limit after invoking a maximum entropy principle to truncate the BBGKY hierarchy.  Earlier motivations for this closure were based on the assumption that each of the pair correlation functions associated with the three edges of the triplet configuration (see Fig. \ref{fig:triplet}) occurs independently of each other \emph{at all spatial scales}, 
 \begin{equation}
  g_3(x_1,x_2,x_3) = g_2 (x_1,x_2) \, g_2 (x_1,x_3) \, g_2 (x_2,x_3).
  \label{eq:kirkDef}
\end{equation}
Substituting the definition of the $k$-th correlation function in terms of the product densities (\ref{eq:kcorrfuncDef}) into (\ref{eq:kirkDef}) for $k=3$ yields a version of the Kirkwood closure (\ref{eq:kirkDef}) that can be used to close the equation at second order (\ref{eq:secondProdDensSpat2}) 
\begin{equation}
   \label{eq:power3}
   \mc(\xi_{1},\xi_{2}) = \frac{
                                \mb(\xi_{1}) \,\mb(\xi_{2}) \,  
                                \mb(\xi_{1}-\xi_{2}) 
                                 }{
                                \ma^3
                               }.
\end{equation}               
This closure also underestimates the second order density, but less dramatically so than the power--2 closure, which results in a slightly better prediction of the mean density (see panel (b) in Figure \ref{fig:compOldClosuresLat}). 
Despite its appealing simplicity,  the power--3 closure shares the same limitations of the other heuristic closures, e.g. there is no criterion of validity, and it provides poor fit to the equilibrium density even for mildly aggregated patterns \cite{rag06} \cite{die00}. 
Heuristic closures have reasonably good performance in random and segregated spatial configurations, but are significantly more limited in aggregated regimes, with the sole exception of the asymmetric power-2. 
  Their limitation arises from the implicit assumption that there are no irreducible triplet correlations at any scale, in the sense that after fixing a pair that forms an edge, for instance the points $x_1$ and $x_2$ (see Fig \ref{fig:triplet}), the two other edges of the triplet formed with the third point $x_3$ occur independently of how the first edge is chosen. This can only be true when the three points are sufficiently far apart, but irreducible third order correlations are likely to occur when the three points are close together in aggregated patterns (Figure \ref{fig:tripletClosure}).  


\subsection{The Maxent closure}
The concept of entropy from an information theoretic point of view,  as opposed to the thermodynamical definition of entropy,  is tightly related to the uncertainty (or information content) associated with an outcome of a random variable.   It can be shown that the information content of a particular outcome $(x'+dx')$ of random variable $x$ with probability density $p(x)$, is given by $\log[p(x')dx']$\cite{sha49,khi57}. The entropy functional is constructed by taking the expected value of the information content over all the possible outcomes of $x$ \cite{sha49,jay57,khi57}.
To illustrate what this means, consider the uniform distribution on an interval $[a,b]\in \R^{+}$. It is not surprising that this distribution maximizes the entropy functional if no constraints are introduced, since all the values in its domain of definition have the same probability weight, thus the uncertainty about a specific outcome of a random variable with this distribution is maximal.  The opposite situation occurs for the Dirac delta distribution which is centered on one single value, say $x'$.  In this situation, a single value occurs with probability one, and all the others have probability zero, therefore the uncertainty about an outcome of this (pathological) random variable is null. \\

The principle of maximum entropy is a powerful method that allows the  derivation of probability distributions when only but a few average properties are all that is known.  Maximizing the entropy functional subject to the constraints provided by these averages, leads to  probability distributions that have the least bias with respect to the known information \cite{jay57,jaynes82,sha49,khi57}.   For instance, maximisation of the entropy constrained to satisfy normalisation and a given mean value leads to the exponential density.  Likewise, maximizing the entropy constrained to satisfy normalisation for a given mean and variance leads to the Gaussian density.  For point processes \cite{fad65,dal04} the entropy is defined with respect to some spatial window of observation $A$, and has two sources of uncertainty, the first is related to the \emph{counts} within $A$, and the second is related to the $\emph{locations}$ of the $n$ points inside this window.  Truncating the hierarchy at order two assumes that only configurations involving up to three points possess irreducible \emph{spatial} information. We carry that assumption forward onto the locational component of the full point process entropy functional, which we then maximise subject to the constraints of normalisation and  product densities up to order two, which are given by the truncated hierarchy.  We exploit formal relationships between the product densities and the probabilistic objects used to construct the entropy functional of a point process ---the \emph{Janossy} densities--- that allow the incorporation of the product density constraints onto the entropy functional, and then translate the results of the maximisation procedure in terms of product densities in order to obtain a closure expression. \\

Our result differs from other maxent closures, like those of Singer \cite{sin04} and Hillen \cite{hil04}, in a number of  ways.  First, it is \emph{implicit}, in the sense that the \emph{third} order density appears in both sides of the closing expression for truncation at second order.  We do so because the Kirkwood closure arises naturally from independence considerations \cite{sin04} for spatial scales larger than the minimum distance for which the pair correlation function is not constant, but it is not valid within the domain of irreducible triplet correlations, i.e. the probability of observing a third point in the triplet depends on how the first two are chosen.  If improvements to the Kirkwood closure are to be made, irreducible triplet correlations must appear in the closure.  In the maxent method we propose irreducible third order correlations are generated by iteration of the closure relationship, while the first and second order densities, generated by the hierarchy,  are held fixed.    Second, we assume that these irreducible third order correlations are confined to a finite window, or spatial scale $A_0$, which is found by comparison of the normalisation condition for the correlated process with that of a Poisson process of the same mean density.  Third, in contrast to other existing approaches, we used all the moments up to the order of the truncation (including the zeroth) to constrain the entropy functional.  This is critically important because the zero-th moment is associated with the normalisation constraint, which allows the determination of the domain of triplet correlations. \\

The variational problem is formulated in terms of the locational entropy functional of the marginal spatial point process.  In order to introduce the product densities as constraints, we exploit known expansions of these in terms of the Janossy densities \cite{dal03,jan50} that constitute the probabilistic objects (the likelihoods) required to construct the entropy functional.  Whereas Singer \cite{sin04}  used the $k$-th order product density to constrain an entropy functional, and Hillen \cite{hil04}, used an $L^{2}$-norm of the moment hierarchy for this purpose, we used instead the classical definition of the entropy functional for a point process, based on the full battery of Janossy densities \cite{fad65,dal04}. \\

The implicit, order two maxent closure (\ref{eq:closurek=2intro}) resembles the structure of the power--3 or Kirkwood closure (\ref{eq:power3}), but is complemented by a number of correction terms that depend on averages of the product densities for each scale at which triplets are irreducible.  Outside this domain, these correction terms vanish and the closure becomes identical to the power--3. There are  two scales of relevance in the closure, one where irreducible triplet correlations are important, and another one where these can be expressed in terms of second and first orders only. \\
  
For the sake of completeness, we first discuss known results related to the entropy of spatial point processes in subsection \ref{ssec:entropy}, and the key expansions of Janossy densities in terms of product densities.  This is followed by the derivation of the implicit maxent closure for truncation at order two (\ref{eq:closurek=2}).  
\subsection{The entropy of a point process}\label{ssec:entropy}
The Shannon (or information) entropy $H[\law]$ of a stochastic process $\law$,  interpreted as the average uncertainty (or information content) associated with a given outcome of $\law$,  is defined as minus the expected value of the log-likelihood $L$ 
\cite{dal03,dal04,jay57,jaynes82,khi57,sha49},
\begin{equation}
 H[\law] = -\E\left\{ \log (L) \right\}.
 \label{eq:entropyDefGeneral}
\end{equation}
The specialisation of the entropy (\ref{eq:entropyDefGeneral}) to point processes  requires a special form of the likelihood, given that in a realisation of a point process of the form $\{x_1,\ldots,x_n\}$ in a window $A$  there are two sources of uncertainty. The first comes from uncertainty about the number of points $n$ within $A$ (the counts), which is controlled by an integer-valued probability distribution $p_n=\Pr \{N(A)=n\}$.  Conditionally on the value of $n$, the other contribution comes from the uncertainty associated with the \emph{locations} of the $n$ points, which is given by a symmetric (in the sense of invariance under permutations of the indices) probability density $s_n(x_1,\ldots,x_n | A)$ on $A^{(n)}$. Thus, the likelihood of a spatial point process is the probability of finding $n$ points within $A$, each in one of the infinitesimal locations $dx_1,\ldots,dx_n$ and nowhere else within $A$. This coincides with the definition of the local Janossy density \cite{dal03,dal04,jan50}
\begin{equation}
  L_A(x_1,\ldots,x_n)=p_n s_n(x_1,\ldots,x_n|A)=j_n(x_1,\ldots,x_n |A).
  \label{eq:definitionLikelihood}
\end{equation}
Separating the contributions due to the counts and those
due to spatial information, we can represent the entropy of a
point process $\N_A$ on a window $A$ by \cite{dal03,dal04}
\begin{equation}
 H[\N_A]=- \sum_{r=0}^{\infty}p_r
 \log(r!p_r)-\sum_{r=1}^{\infty}p_r\int_{A^{(r)}}s_r(x_1,\ldots
 x_r)\,\log[s_r(x_1,\ldots x_r)]\,dx_1\cdots dx_r,
 \label{eq:defEntropyDaley}
\end{equation}
where the integrals calculate the contribution due to the locations, an the sums that of the counts.  If we fix the expected number of points in $A,~\mu=\ma\,|A|=\E[N(A)]$, it can be shown that the first sum in (\ref{eq:defEntropyDaley}) is maximized by the Poisson distribution \cite{dal04,khi57,fad65},
 \[
 p_r=\frac{\mu^{r}}{r!}\exp(-\mu).
 \]
Conditional on the counts $r$, the second sum is maximized by the uniform density on $A^{(r)}$
 \[
 s_r\equiv \frac{1}{|A|^r}.
 \]
Thus, the point process of maximum entropy is the
homogeneous Poisson point process with first order density $\ma$
 \cite{dal08,dal04}.  For closure purposes we  use the definition (\ref{eq:entropyDefGeneral}) written
in terms of the local Janossy densities
\begin{equation}
 H[\N_A]=-\sum_{n=0}^{\infty}\frac{1}{n!}\int_{A^{(n)}}j_n(x_1,
 \ldots,x_n|A)\,\log[j_n(x_1, \ldots,x_n|A)]\,dx_1 \cdots dx_n,
 \label{eq:entropy_functional_janossy}
\end{equation}
where  division by $n!$ ensures normalisation with respect to the $n!$ permutations of the $n$ indices.  Our method of closure consists of maximizing (\ref{eq:entropy_functional_janossy}) constrained to satisfy the product densities up to the order of truncation. These can only be meaningfully  incorporated as constraints if they can be expressed in terms of integrals over $A$ of the Janossy densities.  We do this by using the expansion \cite{dal03},
\begin{equation}
  m_{n}(\xk{n})=\sum_{q=0}^{\infty}\frac{1}{q!}\int_{A^{(q)}}j_{q+n}(\xk{q},y_{1},\ldots,y_  
  {n})\,dy_{1}\dots
  dy_{n}
 \label{eq:facjan},
\end{equation}
where the inverse relationship,
\begin{equation}
 j_{n}(\xk{n}\,|A)=\sum_{q=\,0}^{\infty}
 \frac{(-1)^{q}}{q!}\int_{A^{(q)}}
 m_{n+q}(\xk{n},y_{1},\ldots,y_{q})\,dy_{1}\dots
 dy_{q},
 \label{eq:janfac}
\end{equation}
can be used to translate the results of the constrained optimisation procedure in terms of product densities in order to yield a closure for the product density hierarchy. \\
\subsection{Maximum entropy closure at order $k=2$}\label{ssec:k=2}
\begin{figure} 
\centering
\includegraphics[width=0.7\textwidth]{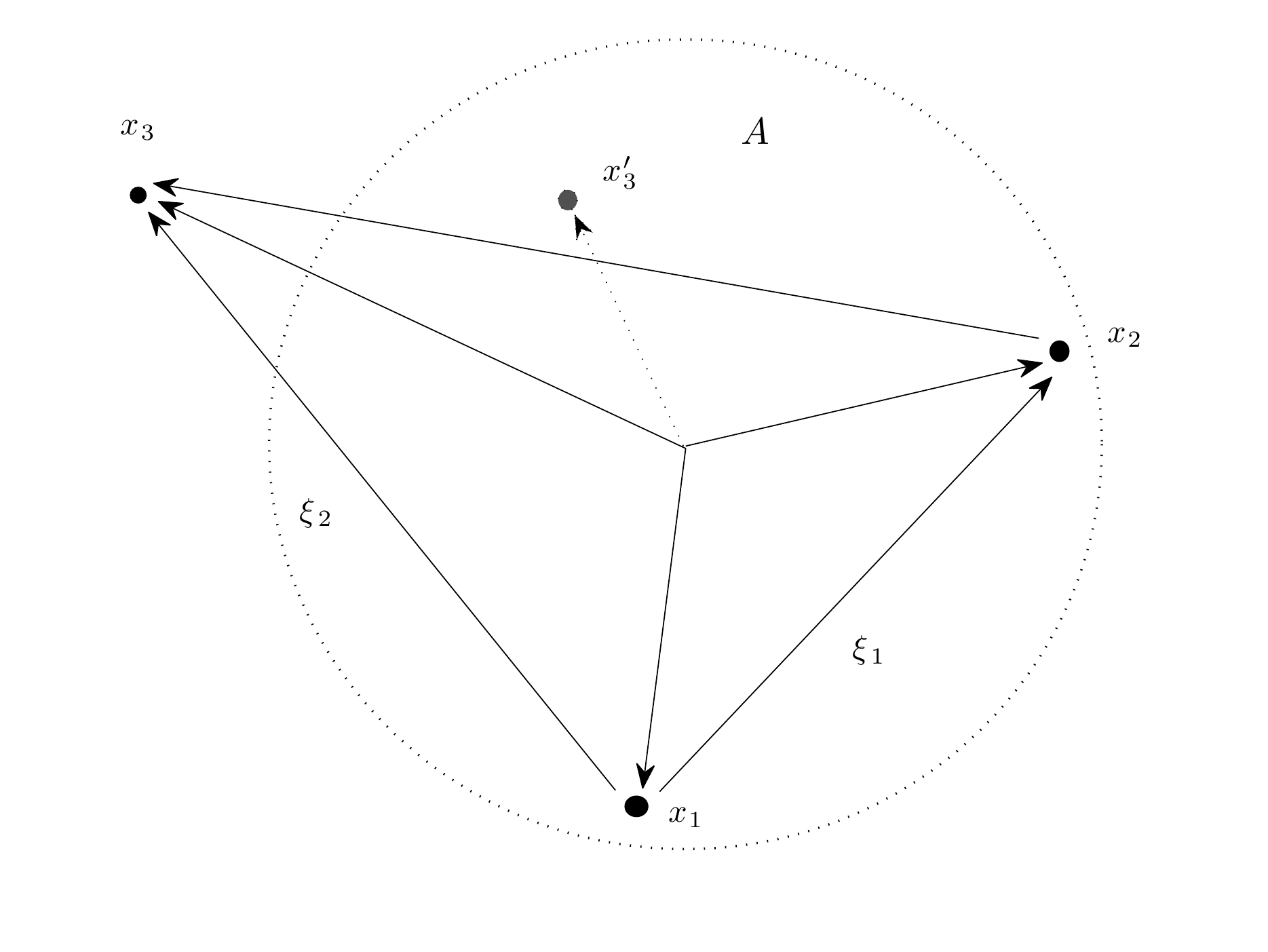}
\caption{
             The domain $A$ represents the region beyond which a third particle becomes independent of the other two.  Shifting $x'_{3}$ to $x_{3}$,  makes that third point independent of the other two, in which case the triplet 
             requires only information about second and first orders density, since the two points along the  $ 
             \xi_{1}$ edge are still correlated. This 
             corresponds to the spatial scale for which the assumptions leading to 
             the Kirkwood closure are valid.
             \label{fig:tripletClosure}
             }
\end{figure}

In the case of the non-homogeneous Poisson point process, which maximizes the entropy functional (\ref{eq:entropy_functional_janossy}), all the points can in principle depend on the specific locations, but these are uncorrelated.  For this special case the expansion of the likelihoods in terms of the product densities (\ref{eq:janfac}) takes the simplified form,
\begin{equation}
 j_{n}(\xk{n}\,|A)=\prod_{p=1}^n \ma(x_p)\sum_{q=\,0}^{\infty}
 \frac{(-1)^{q}}{q!} \prod_{l=0}^q m_{1}(y_l) |A|^l. 
\label{eq:janfacpoisson1}
\end{equation}
If the process is a spatially stationary and homogeneous Poisson point process, then all the product densities become simple powers of the mean density \cite{diggle83,dal03}, which further simplifies (\ref{eq:janfac}) to,
\begin{equation}
 j_{n}(\xk{n}\,|A)=\ma^n \,\exp(-\ma |A|).
\label{eq:janfacpoisson}
\end{equation}
Thus the probability of observing $n$ points within a window $A$ is
\begin{equation}
\Pr \left[N(A)=n\right]=\frac{1}{n!}\int_{A^{(n)}}j_{n}(\xk{n}\,|A)\,\dxk{n},
\label{eq:prnPoisson}
\end{equation}
which after substituting (\ref{eq:janfacpoisson}) into (\ref{eq:prnPoisson}) leads to the Poisson distribution
\[
\Pr \left[N(A)=n\right]=\frac{(m_{1}|A|)^{n}\,\exp(-m_{1}\,|A|)}{n!}.
\]
 \begin{figure} 
  \centering
   \includegraphics[width=0.5\textwidth]{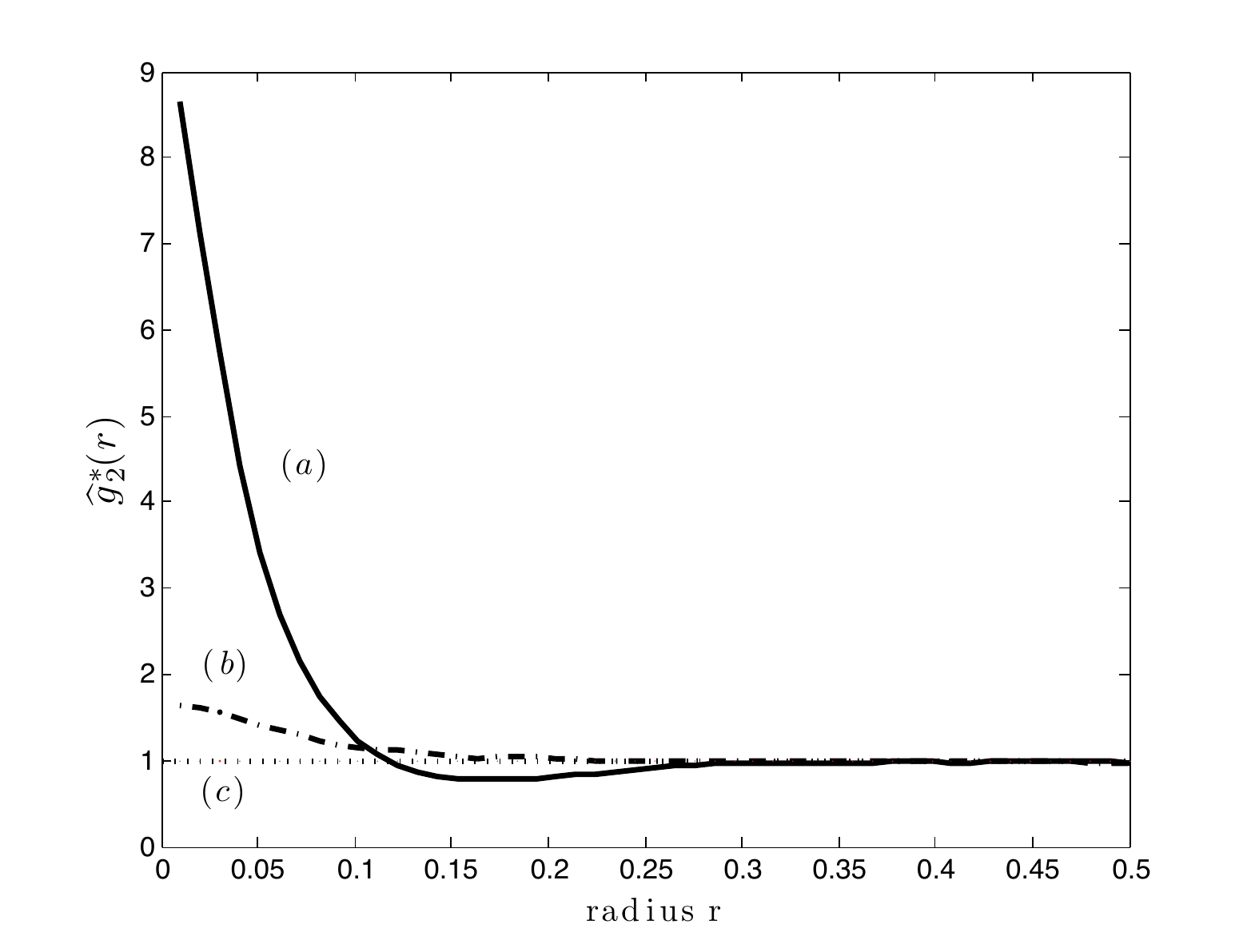}
   \caption{Estimated radial pair correlation functions at
           equilibrium $\hat{g}_{2}^{\ast}(r)$ from  simulations of the point
           process in Section \ref{ssec:simulation_algorithm} with dispersal and mortality 
           kernels given by     
           symmetric bivariate Gaussians.  Parameters lead to a mildly aggregated 
           pattern (case b, dashed line) and a segregated pattern of clusters (case a,  
           continuous line). In (b) we note that correlations decay quickly and become 
           constant at a spatial lag $r > 0.2$, whereas in (a) there are distinct patterns 
           in at least two spatial scales. Aggregation in the smaller ones, and 
           segregation  at intermediate ones. 
           \label{fig:paircorr}
            }
 \end{figure}
We assume somewhat crudely that the Janossy expansions of the point process associated with the moment hierarchy have an intermediate structure between the two extreme cases (\ref{eq:janfac}) where the spatial configurations of all orders are irreducible, and the  Poisson point process (\ref{eq:janfacpoisson}) where all the locations occur independently.  This assumption can be justified from the truncation assumption, since truncating the hierarchy at order two implicitly assumes that terms of order equal or higher than four do not contribute to the formation of second and third order spatial correlations.  Also we see in estimates of the pair correlation functions for the point process discussed in Section \ref{sec:ppmodel}, shown in  Figure (\ref{fig:paircorr}) that there is a region in the parameters for which the spatial correlations of second order decay quickly.  Case (a) corresponds to segregated clusters and thus the pair correlation oscillates around one.  There are two different scales with pattern there.  One associated with the clusters (the region where $g_2>1$) and another with the separation between the clusters themselves ($g_2<1$). Case (b) on the other hand corresponds to  a simply aggregated pattern.  In this latter case we see clearly that there is a spatial scale for which the pair correlation function becomes constant and identical to one, therefore 
\[
m_{2}(r)=m_{1}^{2}, ~~~ r \gg r_{0}
\]
for some spatial scale $r_{0}$.  This assumption is tantamount to requiring that the Janossy expansions of the process to have the form,
\begin{eqnarray}
\nonumber \locj&=&\sum_{q=\,0}^{
k+1-n}\frac{(-1)^q}{q!}\int_{A^{(n)}}m_{n+q}(x_{1},\ldots,x_{n},y_{1},\ldots,y_{q})\,dy_{1}\dots
 dy_{q} \\
 &+&
\prod_{p=1}^{n}\ma(x_{p}) \sum_{q> \,
k+1-n}^{\infty}\frac{(-1)^q}{q!}
\int_{A^{(q)}}\prod_{r=1}^{q}\ma(y_{r})\,dy_r,
 \label{eq:janfacclosable}
\end{eqnarray}
where the first term corresponds to the terms that make contributions due to spatial correlations, and the second term  is the (non-homogeneous) Poisson remainder.  For $k=2$, equation (\ref{eq:janfacclosable}) becomes
\begin{eqnarray}
\nonumber \locj&=&\sum_{q=\,0}^{
3-n}\frac{(-1)^q}{q!}\int_{A^{(n)}}m_{n+q}(x_{1},\ldots,x_{n},y_{1},\ldots,y_{q})\,dy_{1}\dots
 dy_{q} \\
 &+&
\prod_{p=1}^{n}\ma(x_{p}) \sum_{q> \,
3-n}^{\infty}\frac{(-1)^q}{q!}
\int_{A^{(q)}}\prod_{r=1}^{q}\ma(y_{r})\,dy_r.
 \label{eq:janfacclosable}
\end{eqnarray}
The closure assumption implies that only the Janossy densities of order up to $k+1$ make contributions to the \emph{locational} entropy, in which case the entropy functional (\ref{eq:entropy_functional_janossy}) becomes
\begin{eqnarray}
H^{(3)}_{loc}[\N_A] &=&
-J_{0}(A)\log[J_{0}(A)]-\sum_{n=1}^{3}\sum_{1\leq i_{1} <\dots
\leq i_{n} \leq 3}\frac{(3-n)!}{3!}
\\  \nonumber &\times& \int_{A^{(n)}}
j_{n}(x_{i_{1}},\ldots,x_{i_{n}}|A)\,\log[j_{n}(x_{i_{1}},\ldots,x_{i_{n}}|A)]\,dx_{i_{1}}
\cdots dx_{i_{n}}
\label{eq:locentropy3}
\end{eqnarray}
where $J_{0}(A)$ is the avoidance probability in $A$. The first constraint added to (\ref{eq:locentropy3}) is that of normalisation,
\[
1=\sum_{n=0}^{\infty}\frac{1}{n!}\int_{A^{(n)}}j_{n}(\xk{n})\,\dxk{n},
\] 
which after simplification with the assumption (\ref{eq:janfacclosable}) can be added to the entropy functional 
\begin{eqnarray}
\nonumber
+\Lambda_{0}\cdot\left(J_{0}(A)+\sum_{q=1}^{3}\sum_{1\leq i_{1}
 <\dots \leq i_{q} \leq
3}\frac{(3-q)!}{3!}\int_{A^{(q)}}j_{q}(x_{i_{1}},
\ldots,x_{i_{q}}\,|A)\,dx_{i_{1}}\dots dx_{i_{q}} \right.\\
\nonumber \left. +
\sum_{n>3}^{\infty}\prod_{i=1}^{n}\ma(x_{i}) \sum_{l> \,
3-n}^{\infty}\frac{(-1)^l}{l!}
\prod_{r=1}^{l}\int_{A^{(r)}}\ma(y_{r})\,dy_r -1\right)
\end{eqnarray}
where $\Lambda_0$ is a (constant) Lagrange multiplier.  The second constraint is that
of the first order product density $\ma(x_i)$
\begin{eqnarray}
\nonumber &+&\sum_{1\leq i_{1}\leq
3}\frac{1}{3}\int_{A}\Lambda_{1}(x_{i_{1}})\left(\sum_{q=0}^{2}
\sum_{1\leq i_{1} <\dots \leq i_{q} \leq
3}\frac{(3-q)!}{3!}\right.
\\ \nonumber &\times& \int_{A^{(q)}}
j_{1+q}(x_{i_{1}},\ldots,x_{i_{n}},y_{i_{1}},\ldots,y_{i_{q}}\,|A)
dy_{i_{1}}\dots dy_{i_{q}} \\  &-& \left.
\ma(x_{i_{1}})\frac{}{}\right) dx_{i_{1}}.
\label{eq:entrFunctional1m}
\end{eqnarray}
where $\Lambda_1(x_{i_1})$ is a vector of functional Lagrange multipliers, each associated with the permutations in the locations $x_1,x_2$ and $x_3$ comprising the triplet.
Finally, the constraint for the second order product density
$\mb(x_{i_1},x_{i_2})$ is
\begin{eqnarray}
\nonumber &+& \sum_{1\leq i_{1} <i_{2} \leq
3}\frac{1}{6!}\int_{A^{(2)}}\Lambda_{2}(x_{i_{1}},
x_{i_{2}})\left(\sum_{q=0}^{1} \sum_{1\leq i_{1} <\dots \leq i_{q}
\leq 3}\right.
\\ \nonumber && \frac{(3-q)!}{3!}\int_{A^{(q)}}
j_{2+q}(x_{i_{1}},\ldots,x_{i_{n}},y_{i_{1}},\ldots,y_{i_{q}}\,|A)
dy_{i_{1}}\dots dy_{i_{q}} \\  &-& \left.
\mb(x_{i_{1}},x_{i_{2}})\frac{}{}\right)\, dx_{i_{1}}\,
dx_{i_{2}}. 
\label{eq:entrFunctional2m}
\end{eqnarray}
Likewise, the $\Lambda_{2}(x_{i_{1}},
x_{i_{2}})$ are the Lagrange multipliers associated with each of the permutations of the pairs in the triplet. The Euler--Lagrange equations of the functional (\ref{eq:locentropy3})--(\ref{eq:entrFunctional2m}) are
\begin{eqnarray}
\nonumber \frac{\delta H^{(3)}}{\delta J_{0}(A)} =&-&1
-\log[J_{0}(A)] + \Lambda_{0}  = 0,
\\
 \nonumber
\frac{ \delta H^{(3)} 
                             }{
         \delta j_{1}(x_{i_1})
                             } = & - &\frac{1}{3} (1+
\log j_{1}\left[(x_{i_1})\right]) +\frac{1}{3}\Lambda_{0}
+\frac{1}{3}\Lambda_{1}(x_{i_1}) = 0,~~~~~~~~ 1\leq i_1 \leq 3
\\
\nonumber \frac{\delta H^{(3)}}{\delta j_{2}(x_{i_1},x_{i_2})} =&-&
\frac{1}{6}(1+\log \left[j_{2}(x_{i_1},x_{i_2})\right]) +
\frac{1}{6}\Lambda_{0}+\frac{1}{3}\Lambda_{1}(x_{i_1})+\frac{1}{6}\Lambda_{2}(x_{i_1},x_{i_2})
= 0,~~ 1\leq i_1 \leq i_2 \leq 3
\\
\nonumber \frac{\delta H^{(3)}}{\delta j_{3}(x_{1},x_{2},x_{3})}
=& -& \frac{1}{6}(1+\log \left[ j_{3}(x_{1},x_{2},x_{3})\right])
 +
 \frac{1}{6}\Lambda_{0}+\frac{1}{2}\left[\Lambda_{1}(x_{1})+\Lambda_{1}(x_{2})\right.\\
&+&
 \left.\Lambda_{1}(x_{3})\right] +\frac{1}{2}\left[\Lambda_{2}(x_{1},x_{2})+
 \Lambda_{2}(x_{2},x_{3})+\Lambda_{2}(x_{1},x_{3})\right] = 0.
 \label{eq:firstvar}
\end{eqnarray}
It can be seen by inspection that each of the second variations is inversely proportional to minus the Janossy density of order $k$. Since these are all probability densities, each of the second variations is  negative and thus the extrema given in the first variation (\ref{eq:firstvar}) are maxima.
Solving the Euler-Lagrange equations (\ref{eq:firstvar}) for the Lagrange multipliers yields
\begin{eqnarray}
\nonumber \Lambda_{0}&=& 1+\log[J_{0}(A)]\\
\nonumber \Lambda_{1}(x_{1})&=& \log\left[\frac{j_{1}(x_{1})}{J_{0}(A)}\right]\\
\nonumber \Lambda_{1}(x_{2})&=& \log\left[\frac{j_{1}(x_{2})}{J_{0}(A)}\right]\\
\nonumber \Lambda_{1}(x_{3})&=& \log\left[\frac{j_{1}(x_{3})}{J_{0}(A)}\right]\\
\nonumber \Lambda_{2}(x_{1},x_{2})&=& \log\left[\frac{J_{0}(A)\,j_{2}(x_{1},x_{2})}{j_{1}^{2}(x_{1})}\right]\\
\nonumber \Lambda_{2}(x_{2},x_{3})&=& \log\left[\frac{J_{0}(A)\,j_{2}(x_{1},x_{3})}{j_{1}^{2}(x_{2})}\right]\\
 \Lambda_{2}(x_{1},x_{3})&=& \log\left[\frac{J_{0}(A)\,j_{2}(x_{2},x_{3})}{j_{1}^{2}(x_{3})}\right].\label{eq:lagmult}
\end{eqnarray}
After substituting the Lagrange multipliers in (\ref{eq:lagmult}) into the equation for the
first variation with respect to $j_3$ in (\ref{eq:firstvar}) yields an expression that relates the Janossy density of third order to the lower order ones under the assumption of maximum entropy constrained by  the moments, namely 
\begin{equation}
j_{3}(x_{1},x_{2},x_{3}|A)=\frac{j_{2}(x_{1},x_{2}|A)\,j_{2}(x_{2},x_{3}|A)\,j_{2}(x_{1},x_{3}|A)}
{j_{1}(x_{1}|A)\,j_{1}(x_{2}|A\,)j_{1}(x_{3}|A)}\,J_{0}(A),
\label{eq:clojan3}
\end{equation}
Equation (\ref{eq:clojan3}) is formally similar to the Kirkwood closure.  However, there are a number of important differences. First, it varies with the choice of the window $A$, since it depends on the \emph{local} likelihoods (see Figure \ref{fig:tripletClosure}) rather than the product densities used in the Kirkwood closure, which are global properties that do not depend on the window of observation. This  domain $A$ depends on the spatial scale for which the third particle in the triplet becomes independent of the other two.  Second, the closure is weighted by the avoidance probability $J_{0}(A)$.  This term is conceptually similar to the exponential weight suggested by Meeron \cite{meeron57} and Salpeter \cite{salpeter58}, but now arises from a maximum entropy consideration.  The relationship (\ref{eq:clojan3}) can be  used as a closure of the moment hierarchy after using the expansions (\ref{eq:janfac}) and (\ref{eq:janfacclosable}) that allow the Janossy densities to be expressed in terms of product densities. \\
Since the underlying point process is spatially stationary by construction, then the
mean density is constant, and the densities of higher orders depend on the relative rather than absolute distances between points. After rescaling the product densities in the expansion by the area of the window $A$ (the product densities that come from the hierarchy are defined in terms of the much larger spatial window used to observe the full process) we have that the maxent closure is given by \\

if $|\xi_{1}| \leq r_{0}$ and  $|\xi_{2}| \leq r_{0}$ and $ |\xi_{2}-\xi_{1}| \leq r_{0}$
\begin{eqnarray}
\mc(\xi_{1},\xi_{2})\label{eq:closurek=2}
 &=&
\left[\mb(\xi_{1})-\,\ara\int_{A_0}\mc(\xi_{1},\xi_{2}')\,\,d\xi_{2}'\right] \\
\nonumber &\times&
\left[\mb(\xi_{2})-\ara\int_{A_0}\mc(\xi_{2},\xi_{2}-\xi_{1}')\,d\xi_{1}'\right]\\
\nonumber &\times&
\left[\mb(\xi_{2}-\xi_{1})-\,\ara\int_{A_0}\mc(\xi_{2}-\xi_{1}',\xi_{1}')\,d\xi_{1}'\right]\\
\nonumber &\times& \frac{ J_{0}(A_0)}{\left[\ma-\,\ara\int_{A_0}
\mb(\xi_{1}')\,d\xi_{1}'+\frac{\ara^{2}}{2}\int_{A_0^{(2)}}
\mc(\xi_{1}',\xi_{2}')\,d\xi_{1}'\, d\xi_{2}'\right]^{3}},
\end{eqnarray}
else
\begin{equation}
\label{eq:closurek=2kirkout}
  \mc (\xi_{1},\xi_{2}) =
                     \frac{
                     \mb(\xi_{1}) \, \mb(\xi_{2})\,\mb(\xi_{2}-\xi_{1})
                          }{
                          \ma^{3}
                          } 
 \end{equation}
where the circular domain $A_{0}$ of radius $r_{0}$ is determined  from the normalisation constraint (described below).  The avoidance function $J_{0}(A_0)$ is given by 
\begin{eqnarray}
 \nonumber J_{0}(A_0)&=&1-m_{1}\ara+\frac{\ara}{2} \int_{A_0} m_{2}(\xi_{1})d\xi_{1}-\frac{\ara}{6}
 \int_{A_0^{(2)}} m_{3}(\xi_{1},\xi_{2})d\xi_{1}d\xi_{2}  \\
 &+& \sum_{n=4}^{\infty}\frac{(-1)^n}{n!}(m_{1}\ara)^{n}
\label{eq:normalisationk2}
\end{eqnarray}
and the summation term  is equal to
\[
\sum_{n=4}^{\infty}\frac{(-1)^n}{n!}(m_{1}\ara)^{n}= \exp\left(-m_{1}\ara\right)-1+m_{1}|A_0|-\frac{\left(m_{1}|A_0|\right)^{2}}{2}+\frac{\left(m_{1}|A_0|\right)^{3}}{6}.
\]
After simplifying we have
\begin{eqnarray}
J_{0}(A_0)&=&\exp\left(-m_{1}\ara\right) +\frac{\ara}{2}\int_{A_0} m_{2}(\xi_{1})d\xi_{1}-\frac{\left(m_{1}|A_0|\right)^{2}}{2}-\frac{\ara}{6}
 \int_{A_0^{(2)}} m_{3}(\xi_{1},\xi_{2})d\xi_{1}d\xi_{2} \\
 \nonumber &+& \frac{\left(m_{1}|A_0|\right)^{3}}{6}.
\label{eq:normalisationk3}
\end{eqnarray}
In order to obtain the family of sets $A_0$ in the correction terms of the closure, we first need to identify the spatial scale $r_{0}$ beyond which two points become independent. This is equivalent to finding the smallest region  $A_{0}$ for which the correlated point process has the same statistics of a  Poisson process of the same mean density.  This domain is obtained by comparing the avoidance functions for each case, which must coincide for this specific set. Since the avoidance probability for a homogeneous Poisson point process of intensity $m_1$ for some reference window $B$ is equal \cite{dal03} to 
\begin{equation}
\label{eq:avoidanceExp}
J_0^\ast(B)=\exp\left(-m_{1}|B|\right),
\end{equation}
Thus the set $A_{0}$ must satisfy
\begin{equation}
\label{eq:compavoidance}
J_0(A_0)=J_0^\ast(A_0).
\end{equation}
Substituting the rhs of (\ref{eq:avoidanceExp}) and  (\ref{eq:normalisationk3}) into (\ref{eq:compavoidance})  leads to the integral equation 
\begin{equation}
\label{eq:normao}
\int_{A_{r}} m_{2}(\xi_{1})d\xi_{1}-m_{1}^2|A_{r}|-\frac{1}{3}
 \int_{A_{r}^{(2)}} m_{3}(\xi_{1},\xi_{2})d\xi_{1}d\xi_{2} +\frac{
                                                                                               m_{1}^3 |A_{r}|^{2}
                                                                                               }{
                                                                                               3}=0,
\end{equation}
where $A_r=B(0,r)$ is the ball of radius $r$ centered at the origin.  Since all the product densities are given by the hierarchy and the closure relationship (\ref{eq:closurek=2}), the only unknown in (\ref{eq:normao}) is the domain $\ar$ that satisfies the equality (\ref{eq:normao}). This can be found  by evaluating the rhs of (\ref{eq:normao}) for an increasing family of domains $A_r$. The values of for $r$ that satisfy the equality are the roots of interest. There are four possible scenarios for these roots:

\begin{enumerate}
\item The trivial root, $r=0$ is the only solution. This is always a solution by simple inspection.\label{enu:trivial_root}
\item A single non-trivial root $r^\ast$. \label{enu:single_root}
\item A finite number of $n$ non trivial roots $r^\ast_1,r^\ast_2,\ldots,r^\ast_n$. \label{enu:finite_roots}
\item An infinite number of roots.\label{enu:infinite_roots}
\end{enumerate}

A  criterion of validity for the closure scheme can be built on the basis of the number of roots.  Case \ref{enu:trivial_root} indicates that there is not a scale within the observed range of $r$ for which correlations decay as powers of the mean density, and thus truncation should be tried at a higher order. Case \ref{enu:single_root} indicates that there is a single Poisson domain $A_0$ and thus the closure assumptions are consistent with the predicted values of the hierarchy.  Case \ref{enu:finite_roots} indicates that there are several scales of spatial pattern, due to correlations that oscillate as they decay, i.e. segregated clusters (see Figure \ref{fig:paircorr}).  In this situation each scale of pattern should be treated separately.  An infinite number of roots (case \ref{enu:infinite_roots}) indicates that the process is indistinguishable from a Poisson process at all scales. \\

Although the closure expression seems complicated, we note that if the area $a_{0}=|A_{0}|$ is small, 
then the integral correction terms are of similar magnitude, and relatively small in comparison with the correction introduced by avoidance probability,  which by far dominates the closure.  In this situation we have a much simpler approximation to the exact closure, given by
\begin{equation}
\label{eq:simpleclosure}
  \mc (\xi_{1},\xi_{2}) \approx
                     \frac{
                     \mb(\xi_{1}) \, \mb(\xi_{2})\,\mb(\xi_{2}-\xi_{1})
                          }{
                          \ma^{3}
                          } \exp(-\ma \ara).
 \end{equation}
\section{Numerical implementation}\label{sec:numerics}
The numerical solution of the hierarchy with the maxent closure requires two separate modules of code: one  for the integration of the hierarchy itself, and the other for the iterative procedure that computes the third order density.  The first, which we call the `outer' code, consists of a standard numerical integration scheme that predicts  the first and second order product densities at a time $(t+h)$ using the first, second and third order ones at time $t$ as input, where $h$ is a small time step. The second module, or `inner code', computes the third order density at time $(t+h)$ from the maxent closure.  The inner code starts by computing an initial  value for the area of normalisation $A_0^{(old)}$ using the values of the first and second order densities at time $(t+h)$, and the third order density at  time $t$ as an initial trial.  This first value $A_0^{(old)}$ is then substituted in the  maxent closure expression (\ref{eq:closurek=2}) to produce an updated value for the third order density.  The area of normalisation is recalculated with the updated third order density to produce a new value $A_0^{(new)}$; if the relative difference between the old and the new radii associated with each normalisation area  falls below some pre--specified tolerance, then the  iteration stops and the final value of the third order density at time $(t+h)$ is the one being used to calculate the last iteration of area of normalisation.  If not,  the iterations continue until the tolerance is achieved.    We now propose an algorithm for the implementation the maxent closure, and  subsequently show its performance  for a broad range of parameters of the spatial scales. Our
numerical results are well behaved and convergence of the iteration scheme occurs rapidly
for a sufficiently small time step ($h=0.1$), where typically two or three iterations of the closure are sufficient for a relative error tolerance within one percent.
The problem consists of solving the coupled system
  \begin{eqnarray}
      \left\{
          \begin{array}{ccl}
 		\frac{d}{dt}\ma(t) &= &r\, \ma(t)- \dn \int_{\Gamma}
  		 \mor(\xi_1)\,\mb(\xi_1,t)\,d\xi_1 \\
		\\
  		\frac{1}{2}\,\frac{d}{d
 		 t}\mb(\xi_{1},t)&= & b \int_{\Gamma}
   		\dis(\xi_{2})\,\mb(\xi_{1}-\xi_{2},t)\,d\xi_{2}+b\,\dis(\xi_{1})\,\ma(t)-		\di\,
  		 \mb(\xi_{1},t) \\  \\
		 &- &
  		 \dn\, \mor(\xi_{1})\,\mb(\xi_{1},t)-\dn\int_{\Gamma}
   		\mor(\xi_{2})\,\mc(\xi_{1},\xi_{2},t)\,d\xi_{2}.        \end{array}
      \right. 
      \label{eq:momhierarchy}
 \end{eqnarray}
where $\Gamma \subset \R^{2}$ is the computational window.  The  initial condition
\[
 ~~\ma(0)=n_{0},~\mb(\xi_{1},0) = n_{0}^2, ~
      \mc(\xi_{1},\xi_{2},0) = n_{0}^{3}.
\]
The window $\Gamma$ should be large enough to approximate correctly the integral terms so that the scale for which the second and third product densities respectively decay to $m_1^2$ and $m_1^3$ lie well within the computational window$\Gamma$.  This hierarchy can be closed at order 2 with the maxent closure (\ref{eq:closurek=2})
\begin{eqnarray}
\mc(\xi_{1},\xi_{2})\label{eq:closurek=2a}
 &=&
\left[\mb(\xi_{1})-\,\ara\int_{\ar}\mc(\xi_{1},\xi_{2}')\,\,d\xi_{2}'\right] \\
\nonumber &\times&
\left[\mb(\xi_{2})-\ara\int_{\ar}\mc(\xi_{2},\xi_{2}-\xi_{1}')\,d\xi_{1}'\right]\\
\nonumber &\times&
\left[\mb(\xi_{2}-\xi_{1})-\,\ara\int_{\ar}\mc(\xi_{2}-\xi_{1}',\xi_{1}')\,d\xi_{1}'\right]\\
\nonumber &\times& \frac{ J_{0}(\ar)}{\left[\ma-\,\ara\int_{\ar}
\mb(\xi_{1}')\,d\xi_{1}'+\frac{\ara^{2}}{2}\int_{\ar \times \ar}
\mc(\xi_{1}',\xi_{2}')\,d\xi_{1}'\, d\xi_{2}'\right]^{3}},
\end{eqnarray}
which is applied if each the three distance vectors ($\xi_1,\xi_2$ and $\xi_2-\xi_1$, see Figure \ref{fig:tripletClosure}) connecting the three points in the triple  configuration fall  within the normalisation domain $A_{0}$.  Outside of this region we apply the Kirkwood closure on the basis of probabilistic independence of the third point in the triplet, as discussed in the previous section 
\begin{equation}
  \mc (\xi_{1},\xi_{2}) =
                     \frac{
                     \mb(\xi_{1}) \, \mb(\xi_{2})\,\mb(\xi_{2}-\xi_{1})
                          }{
                          \ma^{3}
                          }.
 \end{equation}
In the maxent closure (\ref{eq:closurek=2a}) the avoidance function  $J_{0}(A_0)$ is given by 
\begin{eqnarray}
 \nonumber J_{0}(\ar)=\exp\left(-m_{1}\ara\right).
 \label{eq:normalisationk2a}
\end{eqnarray}
The circular domain $A_{0}$ is computed from the comparison between the normalisation constraint for the truncated hierarchy and that of a Poisson process of the same mean intensity.  It is calculated by finding the value of $r$ that satisfies
\begin{equation}
\label{eq:normaoa}
\int_{A_{r}} m_{2}(\xi_{1}')d\xi_{1}'-m_{1}^2|A_{r}|-\frac{1}{3}
 \int_{A_{r}^{(2)}} m_{3}(\xi_{1}',\xi_{2}')d\xi_{1}'\,d\xi_{2}' +\frac{
                                                                                               m_{1}^3 |A_{r}|^{2}
                                                                                               }{
                                                                                               3}=0.
\end{equation}
where $A_{r}$ is the 2-dimensional ball of radius $r$ centred at the origin. \\

\subsection{Algorithm for the numerical implementation}\label{ref:ssec_num_alg}
The coupled system of product density equations with the maxent closure can be solved from the following algorithm:
\begin{enumerate}
\item \label{enu:radii} From a sequence of radii
$r_i=0,\ldots,r_{max}$, construct an increasing family of  domains $A_{r_{i}}$. \\
 \item At time $t=0$ the initial configuration is given by a homogeneous Poisson point process, thus all the product densities are powers of the mean density $N_{0}/|X|$, where $X$ is the computational spatial arena, and $N_0$ is the population size at time $t=0$. \\
 \item While  the elapsed time $t < T_{max}$ do
 
 \begin{enumerate}
       \item \label{enu:outercode} Integrate forward the densities $\ma(t+h)$ and
               $\mb(\xi_1,t+h)$ from the hierarchy using a standard numerical
             procedure.\\
       \item  Use the value of the triplet density at the earlier time step
              $\mc^{(old)}(\xi_1,\xi_2,t)$ as the initial guess in the normalisation condition for the Poisson  
              area $A_{0}$.  Generate a sequence of values $f(r_{i})$ by calculating the   
              the normalisation condition (\ref{eq:normaoa}) for each  the domains previously constructed in Step \ref{enu:radii} 
              according to
              \begin{eqnarray} 
              \nonumber
               f(r_i)&=&\int_{A_{r_{i}}}\mb(\xi_{1}',t+h)\,d\xi_{1}'-\frac{1}{3}\int_{A_{r_{i}}^{(2)}}
               \mc^{(old)}(\xi_{1}',\xi_{2}',t)\,d\xi_{1}'\,d\xi_{2}'-\ma^{2}(t+h)\,a_{r_{i}}
              \\  &+&\frac{1}{3}\ma^{3}(t+h)\,a_{r_{i}}^{2}
              \end{eqnarray}
               where the $a_{r_{i}}$ are the areas for each of the $A_{r_{i}}$. \\
 	   \item \label{enu:findtheroot} Find the largest
               value $r_{o}$ that satisfies $f(r_o)=0$ by linear interpolation
               between the consecutive $r_{i}$ where $f(r_i)$ changes sign. \\
       \item Use $r_o$ from Step \ref{enu:findtheroot} to generate the estimate of the Poisson
       domain $A_{0}=A_{r_{o}}$.\\
       \item \label{enu:closure} Loop the spatial arguments $\xi_{1}$ and $\xi_{2}$ over the  
       computational spatial arena. \\
       \item Compute the magnitudes
       $d_{1}$, $d_{2}$ and $d_{3}$ of the the distance vectors $\xi_{1}$, $\xi_{2}$ and $\xi_{2}-\xi_{1}$\\
       \item  if $ d_{1} \leq r_{0}$ and $ d_{2} \leq r_{0}$  and $d_{3} \leq r_{0}$ apply the maxent closure\\
  \begin{eqnarray}
         \nonumber
		\mc^{(new)}(\xi_{1},\xi_{2})&=&\frac{\exp(-\ma\,\ara)
		                          }{
		                          \left[
		                            \ma-A_0\int_{A_0}
		                            \mb(\xi_{1}')\,d\xi_{1}'+\frac{A_0^{2}
		                                                       }{
		                                                      2}\int_{A_0}^{(2)}
	                                 \mc^{(old)}(\xi_{1}',\xi_{2}')\,d\xi_{1}'\,d\xi_{2}'\right]^{3}}\\
	       \nonumber & \times & \left[
		                             \mb(\xi_{1})-A_0\int_{A_0}\mc^{(old)}(\xi_{1}',\xi_{2}')\,d\xi_{2}'
	                            \right] 
	                            \\
	                            \nonumber &
		\times & \left[ \mb(\xi_{2})-A_0\int_{A_0}
		\mc^{(old)}(\xi_{2},\xi_{2}-\xi_{1}')\,d\xi_{1}'\right] \\
		\nonumber
		& \times & \left[
		\mb(\xi_{2}-\xi_{1})-A_0\int_{A_0}
	\mc^{(old)}(\xi_{2}-\xi_{1}',\xi_{1}')\,d\xi_{1}'
		\right],\label{enu:closurek=2resc5}
		\end{eqnarray}

      \item else use the Kirkwood closure
        \begin{equation}
        \mc^{(new)}(\xi_1,\xi_2)=\frac{\mb(\xi_{1})\,
        \mb(\xi_{2})\,\mb(\xi_{2}-\xi_{2})}{\ma^{3}}.
        \end{equation}
  
  \item \label{enu:findtherootagain} Recompute the Poisson domain $A_{0}^{(new)}$ and its radius $r_{0}^{(new)}  $ by inserting the corrected triplet density $\mc^{(new)}$ from Step \ref{enu:closure} into the normalisation equation into Step \ref{enu:findtheroot} and estimate a new root $r_n$.\\

\item If the difference between the old radius and the new one falls within the error tolerance
     \[\frac{\left|\,r_o-r_o^{(new)} \right| }{r_{o}
                                    } \leq \mbox{tolerance}
     \] 
     then the third order density at time $(t+h)$ is the one calculated at Step \ref{enu:closure} \\
      \[ 
      \mc(\xi_{1},\xi_{1},t+h)=\mc^{(new)}(\xi_{1}),\xi_{2}
      \]            
     else the old third order density becomes the new third order density
      \[ 
      \mc^{(new)} \rightarrow \mc^{{(old)}}
      \]
       and repeat Steps \ref{enu:findtheroot} through
   \ref{enu:findtherootagain} until the error falls within the tolerance. \\
   \end{enumerate}
   \item update the elapsed time
   \[
     t \rightarrow t +h.
   \]
\end{enumerate}
\subsection{Closure performance}\label{ssec:closureperformance}
We applied the simulation algorithm introduced in the previous Section \ref{ref:ssec_num_alg} using a spatial discretisation of  $47$ points per linear dimension, and the domain $B$ was the  unit square $[-1/2,1/2] \times [-1/2,1/2]$.  The spatial integrals  were computed using the trapezoidal rule, and the convolution in (\ref{eq:momhierarchy}) was calculated using the fast Fourier transform.  For the solution of the moment hierarchy we use a fourth--order Runge-Kutta scheme (with a time step $h=0.1$).  Convergence was checked by halving the time step and the spatial discretisation and no significant differences were found ( $m_1^\star=168.6, \Delta x =1/47,h=0.1$ and $m_1^\star=168.9, \Delta x=1/95, h=0.05$, for $\sigma_W=\sigma_B=0.05$).  \\

\begin{figure*}
\centering
\includegraphics[width=\textwidth]{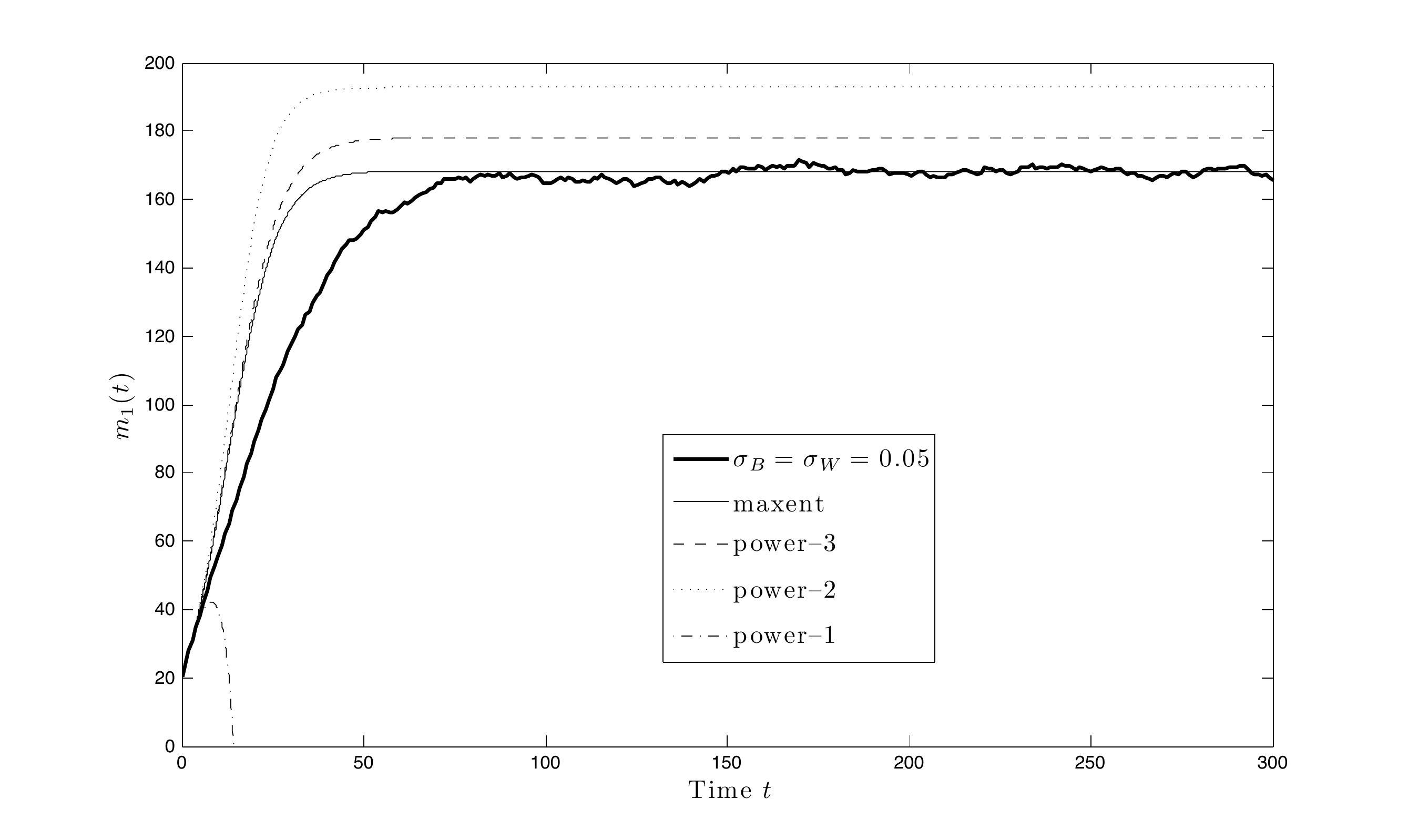}
\caption{Comparison between the mean density (jagged blue line)
for a sample of 300 simulations of the point process for the
mildly aggregated case $\sigma_{\dis}=0.05, \sigma_{\mor}=0.05$ (open circles) and the truncated product density hierarchy using various closures. The 
the maximum entropy closure (maxent) (continuous black line), the
power--3 (dash-dot), the symmetric power--2 (dot) and
power--1 (dashed). The maximum entropy closure provides the best
fit to the equilibrium values of the IBM. However the performance of all the closures is poor during the transient regime.}
\label{fig:compClosures0505}
\end{figure*}

The maxent closure is expected work well in situations where the spatial scales of dispersal and mortality are similar, since this combination of parameters tends to produce a single scale of spatial pattern of mild aggregation (see Figure \ref{fig:paircorr}), where higher order terms are small.  Figure \ref{fig:compClosures0505}  compares the dynamics of the mean density predicted by the maxent closure in a mildly aggregated regime ($\sigma_\dis=\sigma_\mor=0.05$ ) against averages of the point process model and the other closure methods used in the literature, power--1, power--2 and power--3 (but the \emph{asymmetric} power--2 is not used in the comparison).  We see that the maxent closure outperforms the other closures. As before, in all cases the transient is predicted poorly.  This is to be expected of the maxent method, because the locational entropy can be assumed to be maximised only once the stochastic process has reached its stationary distribution.  For this reason, even with  the correction terms, the truncated hierarchy with the maxent closure fails at capturing the transient behavior, which typically consists of long range spatial correlations that decay only once the density--dependent mortality term is large enough to cause mixing at longer scales, thus producing a shorter correlation scale.\\
%

\begin{figure*}
\centering
\includegraphics[width=\textwidth]{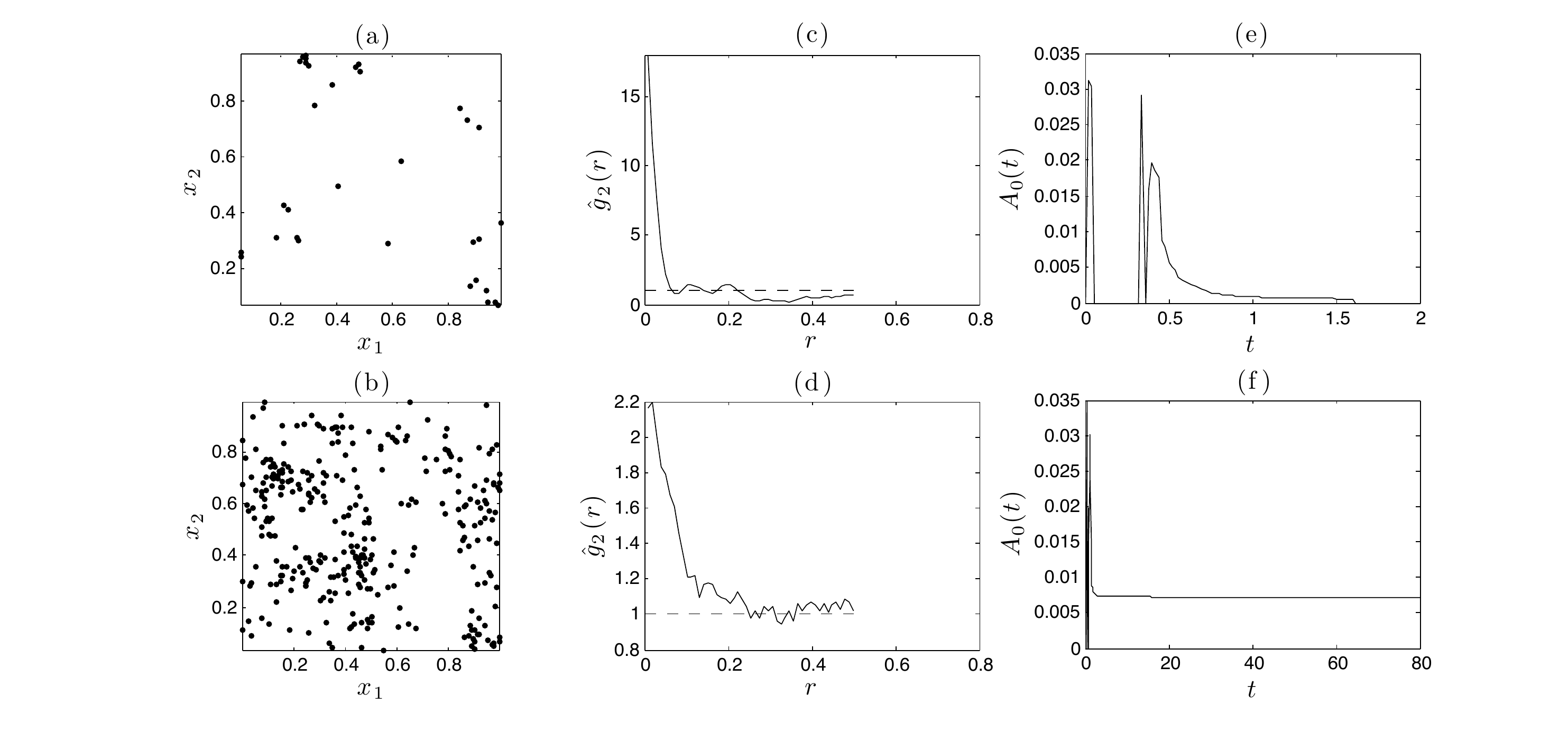}
\caption{
Behavior of the area of corrections in the maxent closure for two types of agregated spatial patterns.  The upper three panels correspond to a segregated pattern of clusters with $\sigma_\dis=0.02,\,\sigma_\mor=0.12$, and the lower panels to a mildly aggregated pattern with $\sigma_{\dis}=\sigma_{\mor}=0.04$. The left column shows a single point pattern at the end of the simulation, the middle column shows a kernel density estimate of the pair correlation function for the pattern displayed in the left and the right column shows the temporal behavior of the area of the set in the correction terms.  
}\label{fig:compb_and_dpatterng2andarea1}
\end{figure*}
The ability of the maxent closure to predict accurately the mean density changes dramatically when the two interaction kernels have very different characteristic scales. This combination of parameters leads to several scales of pattern, that can consist of short range aggregation compensated by long range segregation, or short scale segregation compensated by long range clustering.  This occurs because the total number of pairs over sufficiently long ranges must be equal to the density squared.  Thus, extreme aggregation over short scales must  be compensated by segregation over the longer scales in order to preserve the total number of pairs.  When dispersal has a much shorter characteristic scale than that of density--dependent mortality, the resulting pattern consists of segregated clusters.  This situation violates the closure assumptions (that require a single scale of pattern), and we expect the validity checks in the maxent  closure to be activated in this situation.   This is illustrated for two types of aggregated patterns in Figure \ref{fig:compb_and_dpatterng2andarea1}.   The upper three panels correspond to segregated clusters ($\sigma_\dis=0.02,\sigma_\mor=0.12$), and the lower three to the mild aggregation case discussed earlier ($\sigma_\dis=\sigma_\mor=0.04$). The left column conformed by panels (a) and (b) show typical point patterns obtained at the same time at which the numerical solution of the hierarchy stopped, $t=1.56$ in (a), because of the validity check, and $t=80$ in (b) which was long enough to reach equilibirium.  The center column, consisting of panels (c) and (d), displays  kernel density estimates of the pair correlation function for the point patterns shown to the left. We see in panel (c) a very high degree of aggregation at short scales followed by long range segregation. Finally,  panels (e) and (f) show the dynamics of the area of correlations $A_{0}(t)$ for both regimes.  We see  failure of the maxent closure to find a non-trivial root for $A_{0}$ in panel (e) after a short transient, as should be expected due to the presence of various scales of pattern detected in the pair correlation function in panel (c). In this situation, the extreme form of  `checkerboard' aggregation requires truncation at a higher order. Since the pair correlation function is clearly not constant, but yet the normalisation constraint only finds the trivial root zero, the validity check is activated and the numerical solution of the hierarchy stops.  By contrast, in the lower panels when the degree of clustering is comparatively smaller, the method succeeds in finding a single root $A_{0}$ that eventually reaches a single equilibrium (see panel (f)).   \\

\begin{figure*}
\centering
\includegraphics[width=1.1\textwidth]{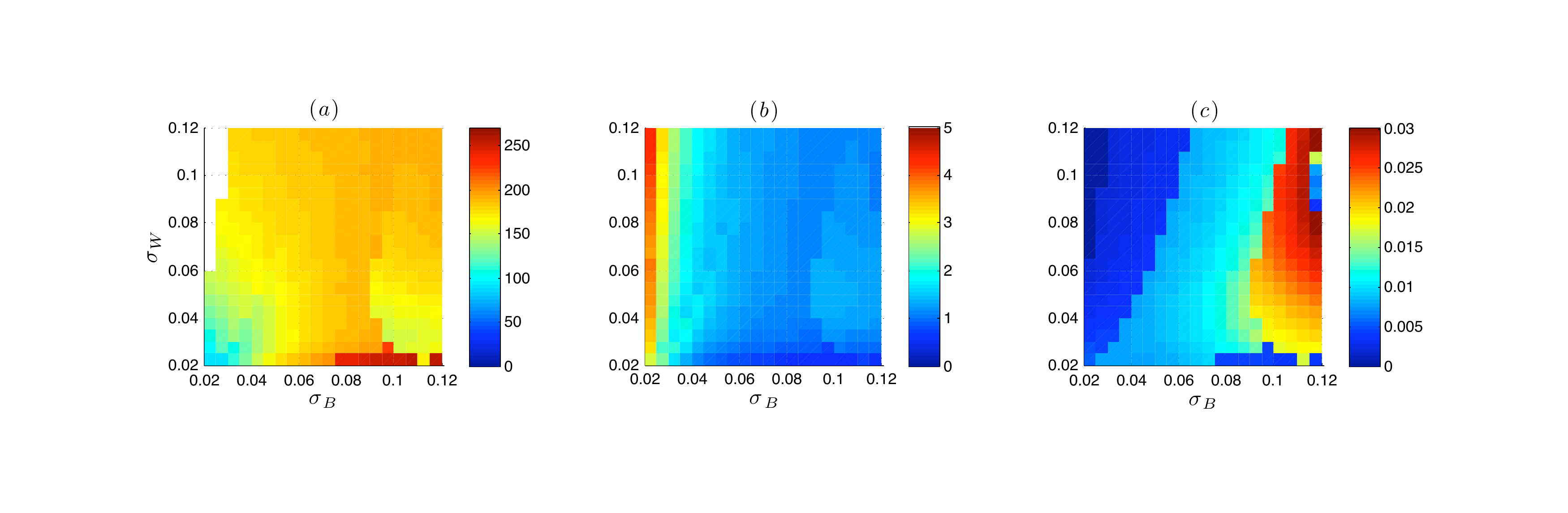}
\caption{Simulation results of the product density hierarchy with the maxent for various values of the characteristic spatial scales of dispersal $\sigma_{B}$ (horizontal axis) and mortality $\sigma_{W}$(vertical axis).  The left panel (a) shows the equilibrium mean density $m_{1^{\ast}}$. The center panel shows the value of the second order product density at equilibrium evaluated at the origin, normalized by the squared mean density.  In this panel values higher than one  indicate clustering at short scales, and values below one indicate segregation.  The right panel (c) shows the value at equilibrium of the area of the domain used in the correction terms $A_{0}$.}
\label{fig:maxent47stata}
\end{figure*}

We carried out a systematic exploration of the behavior of the maxent closure for a wide range of  combinations (441 in total) of the spatial parameters falling within the range $[0.02, 0.12]$ that correspond to those explored earlier by Law \emph{et al}  \cite{law03}, and compare the results with the predictions of the point process, and the product density hierarchy with the power--3 closure. This  allows the assessment of the relative importance of the correction terms in the maxent closure. The upper limit in the parameter domain was chosen because for that scale ($\sigma_{B}=\sigma_{W}=0.12$) there is only a very small departure from complete spatial randomness. Figure \ref{fig:maxent47stata} shows various equilibrium values predicted by the product density hierarchy with the maxent closure.  Panel (a) corresponds to the mean density, panel (b) shows the equilibrium value of the second moment at the origin, normalized by the mean density squared, and finally, panel (c) shows the area of normalisation at equilibrium.  The removed regions (white) in panel (a) result from the application of the validity check of normalisation, since for this parameter the area of correlations is zero (see panel (c)), but the second order product density indicates the existence of spatial pattern.  \\

\begin{figure*}
\centering
\includegraphics[width=1.1\textwidth]{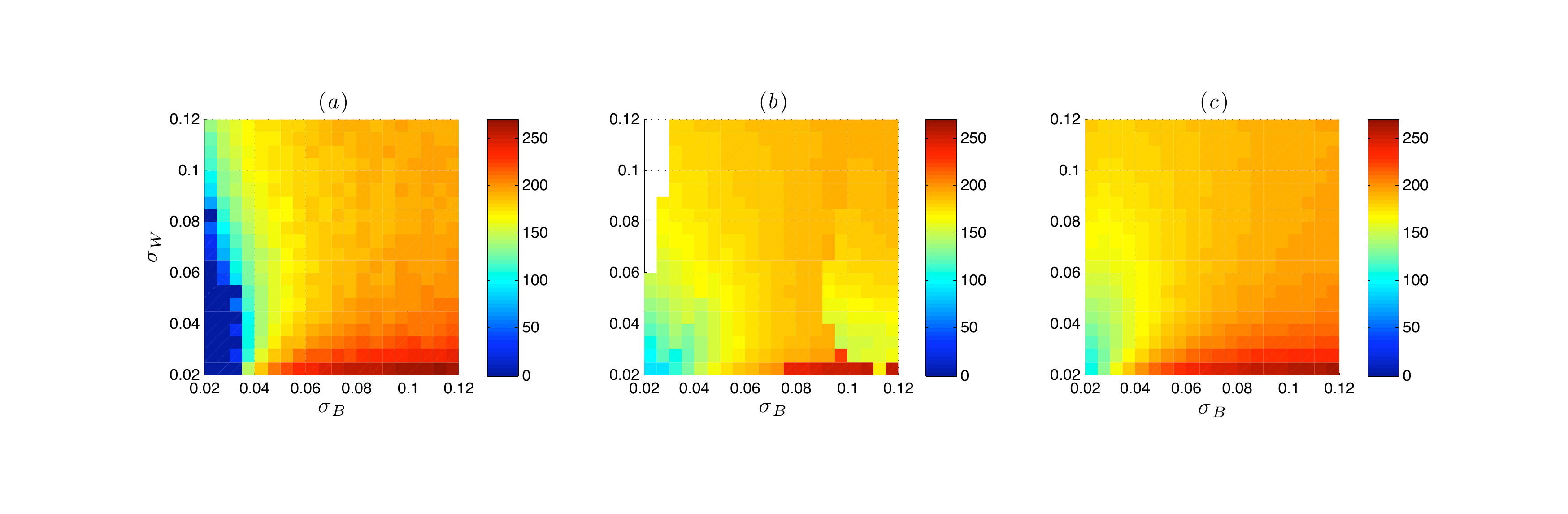}
\caption{Comparision of the mean density $m_{1}^{\ast}$ at equilibrium predicted by an ensemble average of the point process model (a),  the maxent closure (b), and the power--3 or Kirkwood closure (c).  In panel (b) the white region no the upper left corner corresponds to the domain where the normalisation constraint returns a trivial root for values of the second order product density that indicate the presence of spatial pattern, activating the validity check (\ref{eq:normao}).}
\label{fig:m1pp_maxent_kirka}
\end{figure*}

%
In Figure \ref{fig:m1pp_maxent_kirka} we compare the mean equilibrium density predicted from an average of the the space--time point process (a), the maxent closure (b), and the power--3 closure (c). The maxent closure is not a good predictor of the mean density for intermediate to low ranges of mortality combined with long range dispersal;  in this regime both the qualitative and quantitative behavior of the closure is poor. We see a sharp drop in the values of the mean density, whereas in the point process model it grows monotonically before reaching the plateau that  occurs when both dispersal and mortality act over long scales.  This combination of parameters leads to segregation at short scales and long range (albeit mild) aggregation.  The maxent method detects only the scale of aggregation, which produces comparatively larger values of the area of correlations (see panel (c) in Figure \ref{fig:maxent47stata}).  This leads to over--correction in the maxent closure, which results in an equilibrium density that falls well below that predicted by the point process model.  In this regime, the power--3 closure provides a much more precise prediction of the equilibrium density, both qualitatively and quantitatively. 
For sufficiently short ranges of dispersal together with short to intermediate ranges of mortality  the point process model predicts extinction, as already noted earlier by \cite{law00,law03}.  In this regime, neither the maxent closure nor the power---3 closure is capable of predicting the persistance/extinction threshold, and the maxent validity check does not seem to operate either.  However,  for intermediate ranges of aggregation close or above the main diagonal ($\sigma_W=\sigma_B$),  the maxent closure does provide an improved prediction of the equilibrium density, with the added benefit of the criterion of validity being activated when dispersal is short range with long range mortality, which leads to different scales of pattern. \\

\begin{figure}
\centering
\includegraphics[width=0.5\textwidth]{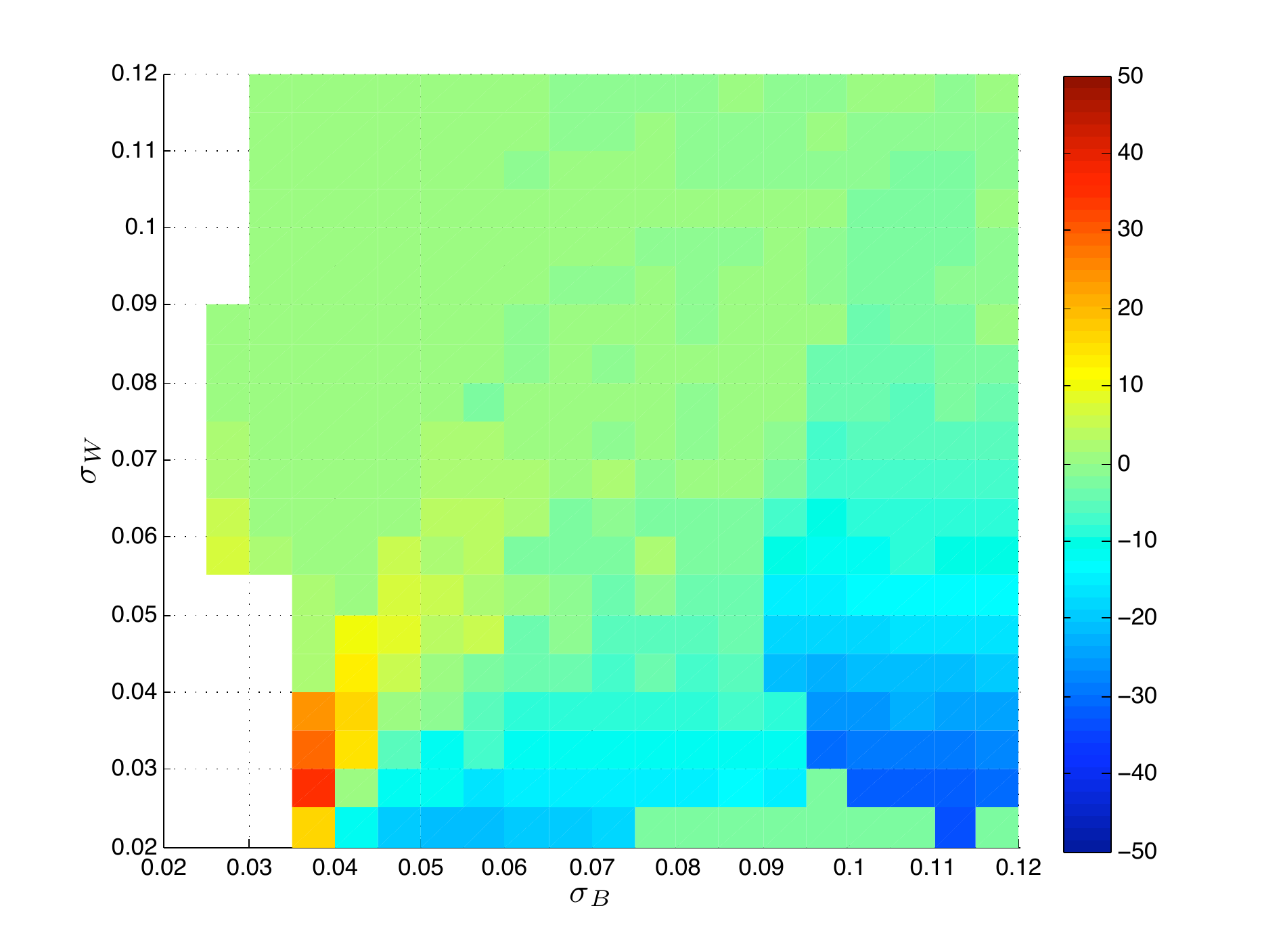}
\caption{
Difference in relative error between the maxent and power--3 closures for various combinations of dispersal and mortality spatial scales.  Values higher than zero indicate that the maxent closure outperforms the power--3 closure, whereas negative values are evidence of better precision of the power--3 closure.}
\label{fig:errorcompa}
\end{figure}

We computed the relative error between the equilibrium density of the point process, and that predicted by the moment equations with the two closures, shown in Figure \ref{fig:m1pp_maxent_kirka}. Panel (a) corresponds to the maxent closure and panel (b) to the power--3.  We see that the maxent closure has larger  relative error than the power--3 for values located below the diagonal ($\sigma_W=\sigma_B$), which are associated with segregated spatial patterns (see Figure \ref{fig:maxent47stata}, panel (b)).  In contrast, the power--3 closure performs quite well in this region.  The advantage of the  maxent closure becomes more noticeable on, and above the diagonal, which is associated with aggregated patterns. The ability to predict correctly the equilibrium density in this regime is nearly optimal; particularly when the two scales have similar magnitudes, even when both dispersal and mortality act over short ranges.  The regions of the parameter space for which each of the two closures is relatively more useful are shown in Figure \ref{fig:errorcompa},  which displays the difference in relative error between the two closures $\Delta E = err_{p3}-err_{maxent}$. Positive values of $\Delta E$ indicate that the error in the power-3 closure is larger than the maxent closure, and vice versa for negative values of $\Delta E$.  As discussed above the largest improvement of the maxent closure  around to the region where the two scales are of similar magnitude.
\begin{figure*}
\centering
\includegraphics[width=1.1\textwidth]{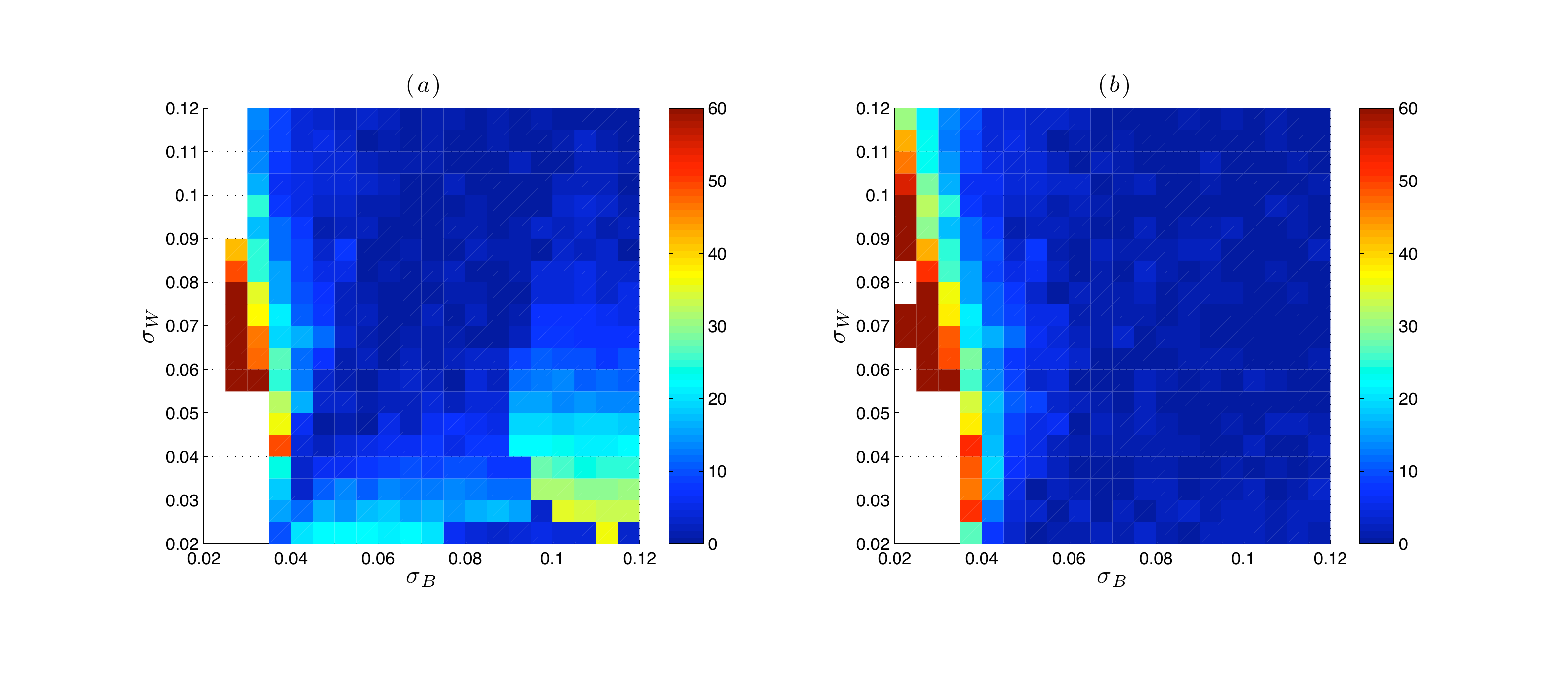}
\caption{Relative error of the maxent closure (a) and the power--3 closure (b).  We see that the maxent closure performs better than the power--three closure for mildly aggregated patterns (lower left), but the Kirkwood closure outperforms the maxent in segregated patterns (lower right)}
\label{fig:relativeerrormaxentkirka}
\end{figure*}



\section{Discussion}
\label{sec:conclusions}
The results of this research resonate with previous work \cite{bol97,law03,ovaskainen06} that demonstrates that the analysis of stochastic, locally-regulated, individual-based  models of population dynamics in continuous space is feasible \cite{ovaskainen06,bol97,law03}.  The numerical implementation of the maxent closure is computationally more expensive (about twice as much) than existing closure methods, but is nonetheless faster than resorting to direct simulation of the point process; if one is willing to approximate, the simplified closure based solely on the exponential correction (\ref{eq:simpleclosure}) is substantially simpler to implement, and produces very small errors in comparison with the full maxent closure. Although a number of moment closures have been proposed in the literature, some using heuristic arguments, and others based on constrained entropy maximisation, very few, if any have a criterion of validity, with the  exception of Ovaskainen \& Cornell \cite{ovaskainen06} who were able to derive a series expansion for the mean density of a spatially explicit metapopulation problem, and show rigorously that their approximation to the mean density is exact in the limit of long range interactions.  The principal benefit of the maxent method  lies in the fact that the normalisation constraint used to find the domain for the correction terms fails to find a non-trivial root when the closure assumptions are not met.  This situation occurs when higher order terms are required in order to fully capture the dynamics, or when correlations extend over a range that goes beyond the window of observation. This property constitutes a validation check, not present in other proposed closure schemes. \\

Although the power--3 or Kirkwood closure had previously been  derived from maximum entropy arguments \cite{sin04} (but using a different set of constraints and a different definition of the entropy functional), the correction terms presented here are new, and extend the probabilistic interpretation of the Kirkwood closure to situations where there is a region of irreducible triplet correlations.  These correction terms introduce significant improvements in the agreement between the simulations of the stochastic process (for mildly aggregated patterns) and its deterministic approximation by the product density hierarchy. It remains to be seen how the maxent closure behaves for other functional forms of the interaction kernels, particularly for those that have tails that decay algebraically ( i.e. power laws) instead of exponential. Another area of further work would be related to changes in the value of the non--spatial carrying capacity $K$.  For higher densities, spatial effects become less important.  \\

Since the derivation of the method does not depend on the details of the model,  but only on that its equilibrium distribution is of maximum \emph{locational} entropy with moment constraints, the maxent closure may be useful beyond spatial ecology where unclosed hierarchies for particle distribution functions are also commonly found, for instance in the statistical mechanics of fluids where the Kirkwood closure was first introduced \cite{sin04}, or in problems where the organisms move in space \cite{bir06,fli99,you01}, provided that the correlation functions in those models are stationary in both space and time.  A limitation of the method is its poor ability to predict the transient. This is to be expected, since maximum entropy is a meaningful property of the \emph{equilibrium} distribution only when detailed balance is satisfied \cite{van01,gar85,jay57} and the transitions due to fecundity and dispersal events coincide with mortality.
Other areas of current and future work include the generalisation of the moment hierarchy and the maxent closure to an arbitrary order of truncation, extensions to \emph{marked} spatial point processes for populations with both spatial and size structure. \\



\appendix{Appendix.  Derivation of moment equations}
 \begin{appendix}
 In order to derive the equation for $\ma(t)$, we start by
fixing a small region of observation $dx_1$ (so that the count inside $dx_1, N(dx_{1})$ is either 0 or 1) and write a Master equation for 
the probabilities of change in the count $\Delta N_{\delta t}(dx_1)$
during a small time interval $\delta t$, defined as
\[
\Delta N_{\delta t}(dx_1)=N_{t+\delta t}(dx_1)-N_t(dx_1).
\]
These come from the birth and death transitions. Births are given
by the probability that the count $N(dx_1)$ increases by one in
$\delta t$ due to a birth in $dx_1$
\[
 N \mapsto N+1,
\]
This probability is controlled by the fecundity rate and the dispersal kernel,
\begin{eqnarray}
\nonumber f(x_1| \varphi_{t} )&=& \pr \left\{ \mbox{ one birth in
} (dx_1) \mbox{
 during } (t,t+\delta t) \, |\, \varphi_{t}(X) \right\}.\\
&=& \left[b \sum_{x_{n} \in \varphi_{t}}
\dis(x_1-x_{n})\,N_t(dx_n)\ell(dx_1)\right]\delta t + o(\delta t),
\label{eq:brate}
\end{eqnarray}
where $b$ is  the birth rate, $\dis(\xi)$ is the dispersal kernel,
$\varphi_{t}$ is the configuration of points at time $t$ and $\ell(A)$ is the area of the 2-dimensional domain $A$. For the death
of the individual in $dx_1$, we have the transition
\[
 N \mapsto N-1,
\]
controlled by
\begin{eqnarray}
\mu (x_1| \varphi_{t} )&=& \pr \left\{ \mbox{ death of individual
} x_1 \mbox{ during } (t,t+\delta t)\, |\, \varphi_{t} (X)
\right\}.\\ \nonumber &=& N_t(dx_1)\left[d+ \dn \sum_{x_{n}\in
\varphi_{t}}
\mor(x_1-x_{n})\left(N_t(dx_n)-\delta_{x_1}(dx_n)\right)\right]\delta
t + o(\delta t) \label{eq:drate},
\end{eqnarray}
where $d$ and $\dn$ are positive constants defined in Section \ref{sec:ppmodel}, the
density--independent, and density--dependent contributions to the
mortality and $\mor(\xi)$ is the mortality kernel).  This probability is conditional on
there being an individual in $dx_1$. The change in the count $\Delta N_{\delta
t}(x_1)$ is then given by both contributions
\[
\Delta N_{\delta t}(dx_1)=f(x_1|\varphi_t)-\mu(x_1|\varphi_t)
\]
so
\begin{eqnarray}
\Delta N_{\delta t}(dx_1)&=&\left[b \sum_{x_{n} \in
\varphi_{t}} B(x_1-x_{n})\,N_{t}(dx_n)\,\ell(dx_1) \right. \label{eq:transDN1} \\
\nonumber &-& \left. N_t(dx_1)\left(d+ \dn \sum_{x_{n} \in
\varphi_{t}}
W(x_1-x_{n})(N_{t}(dx_n)-\delta_{x_1}(dx_n)\,\right)\right]\delta
t.
\end{eqnarray}
Taking expectations (ensemble averaging) on both sides and dividing by the duration of a small time interval $\delta t$
yields
\begin{eqnarray}
\nonumber \frac{\E\{\Delta N_{\delta t}(dx_1)\}}{\delta t}&=&b
\sum_{x_{n} \in
\varphi_{t}} \dis(x_1-x_{n})\,\E\{N_{t}(dx_n)\}\,\ell(dx_1) \\
\nonumber &-& \E\left\{N_t(dx_1)\left(d+ \dn \sum_{x_{n}\in
\varphi_{t}}
\mor(x_1-x_{n})(N_{t}(dx_n)-\delta_{x_1}(dx_n))\right)\right\}.
\end{eqnarray}
after rearranging the second term, dividing both sides by
$\ell(dx_1)$ and multiplying the second sum by $\ell(dx_n)/
\ell(dx_n)$ we get
\begin{eqnarray}
\nonumber \frac{\E\{\Delta N_{\delta
t}(dx_1)\}}{\ell(dx_1)\,\delta t}&=&b
\frac{\E\{N_{t}(dx_n)\}}{\ell(dx_1)} \sum_{x_{n} \in \varphi_{t}}
B(x_1-x_{n})\,\ell(dx_1) - d\,\frac{\E\{N_t(dx_1)\}}{\ell(dx_1)}
\\ \nonumber
 &-& \dn\sum_{x_{n}\in \varphi_{t}}W(x_1-x_{n})\frac{\E \left\{N_{t}(dx_1)\,\left( N_{t}(dx_n)-
 \delta_{x_1}(dx_n)\right)\right\}}{\ell(dx_{1})\,\ell(dx_n)}\,\ell(dx_n).
\end{eqnarray}
taking the limits as $\ell(dx_1)$ and $\ell(dx_n)$ go to zero, and
using definition of the product density (\ref{eq:genDefProdDens})
\begin{eqnarray}
\nonumber \frac{\Delta \ma(x_1,t)}{\delta t}&=&b\,
\ma(x_{1},t)\int_{\Re^2} \dis(x_1-x_{n})\,dx_1 -
d\,\ma(x_{1},t)
\\ \nonumber
 &-& \dn\int_{\Re^2} \mor(x_1-x_{n})\,\mb(x_{1},x_{n},t)\,dx_n.
\end{eqnarray}
since the process is  spatially  stationary by construction and exploiting the
fact that the dispersal kernel integrates to unity, yields
\begin{eqnarray}
\nonumber \frac{\Delta \ma(t)}{\delta t}=b\, \ma(t)-
d\,\ma(t)- \dn\int_{\Re^2} \mor(\xi_1)\,\mb(\xi_1,t)\,d\xi_1,
\end{eqnarray}
finally, after taking the limit as $\delta t \rightarrow 0$ we
get,
\begin{eqnarray}
\frac{d}{dt}\ma(t)=b\, \ma(t)- d\,\ma(t)-
\dn\int_{\Re^2} \mor(\xi_1)\,\mb(\xi_1,t)\,d\xi_1.
\end{eqnarray}
On setting $r=b-d$, we get the the generalisation of the
logistic equation to the spatial case obtained by  Law \& Dieckmann \cite{law00} and Law
\emph{et al,law03}, but derived explicitly in terms of
product densities,
\begin{eqnarray}
\frac{d}{dt}\ma(t)=r\, \ma(t)- \dn\int_{\Re^2}
\mor(\xi_1)\,\mb(\xi_1,t)\,d\xi_1 \label{eq:firstProdDensSpatr}.
\end{eqnarray}
Since $\mb$ is unknown, we need an
additional evolution equation for this object.  We follow a similar procedure to that used for the mean density, but considering the expected change of the \emph{product} of the counts in two observation regions $dx_1$ and $dx_2$. This requires the consideration of how pairs of points are created and destroyed as individuals disperse and die. There are three
possible ways in which changes to occur. The first if to fix the count
$N_{t}(dx_1)$ and allow only $N_{t}(dx_2)$ to change. The second
is the reverse situation, fixing $N_{t}(dx_2)$ and allowing
only $N_{t}(dx_1)$ to change.  The third is when \emph{both}
$N_t(dx_1)$ and $N_t(dx_2)$ change in a small time interval.  We
have that
\begin{eqnarray}
\Delta[N_{t}(dx_1)\,(N_{t}(dx_2)-\delta_{x_1}(dx_2))]&=&
N_{t}(dx_1)\Delta(N_{t}(dx_2)-\delta_{x_1}(dx_2))
\\ \nonumber &+& (N_{t}(dx_2)-\delta_{x_1}(dx_2))\Delta N_{t}(dx_1) \\ \nonumber
&+& \Delta N_{t}(dx_1)\Delta (N_{t}(dx_2)-\delta_{x_1}(dx_2))
\end{eqnarray}
where the Dirac delta distribution is used to remove self-pairs. The following derivation for the second order product densities is based on the symmetry in the probabilities of a birth or a death event occurring at both extremes of the distance vector linking $x_1$ and $x_2$.
We
also assume that a \emph{simultaneous} change in both
$N_{t}(dx_2)$ and $N_{t}(dx_1)$ is negligible
\[
\pr \left[\Delta N_{t}(dx_1)\Delta(N_{t}(dx_2)-\delta_{x_1}(dx_2)) \right]=o(\delta
t)
\]
and thus the transitions of second
order can be written as
\begin{equation}
\Delta[N_{t}(dx_1)\,(N_{t}(dx_2)-\delta_{x_1}(dx_2))]=2\Delta
N_{t}(dx_1)(N_{t}(dx_2)-\delta_{x_1}(dx_2)).
\label{eq:transNNHeur}
\end{equation}
Since we already have an expression for $\Delta N_t(dx_1)$, given
by  (\ref{eq:transDN1}), (\ref{eq:transNNHeur}) becomes
\begin{eqnarray}
\nonumber
\Delta[N_{t}(dx_1)\,(N_{t}(dx_2)-\delta_{x_1}(dx_2))]&=&2 \cdot
(N_{t}(dx_2)-\delta_{x_1}(dx_2))\left[b \sum_{x_{n} \in
\varphi_{t}} B(x_1-x_{n})\,N_{t}(dx_n)\,\ell(dx_1) \right. \\
\nonumber &-& \left. N_t(dx_1)\left(d+ \dn \sum_{x_{n} \in
\varphi_{t}}
W(x_1-x_{n})(N_{t}(dx_n)-\delta_{x_1}(dx_n)\,\right)\right]\delta
t. \label{eq:transDN2}
\end{eqnarray}
Taking expectations, and dividing by both sides by $\delta t$
gives
\begin{eqnarray}
\nonumber
\frac{\Delta[N_{t}(dx_1)\,(N_{t}(dx_2)-\delta_{x_1}(dx_2))]}{2\,\delta
t}&=&b \sum_{x_{n} \in \varphi_{t}} B(x_1-x_{n})\,\E\left
\{N_{t}(dx_n)(N_{t}(dx_2) \right.\\ \nonumber &-&
\left.\delta_{x_1}(dx_2))\right.\}\,\ell(dx_1)
-d\,\E\left\{N_t(dx_1)(N_{t}(dx_2)-\delta_{x_1}(dx_2))\right\}\\
\nonumber &-& \dn\, \sum_{x_{n}\in \varphi_{t}}
W(x_1-x_{n})\E\left\{N_t(dx_1)(N_{t}(dx_n) \right.
\\ \nonumber &-&
\left.\delta_{x_1}(dx_n))(N_{t}(dx_2)-\delta_{x_1}(dx_2))\right\}.
\end{eqnarray}
After dividing by $\ell(dx_1)$ and $\ell(dx_2)$, using the
definition of product densities (\ref{eq:genDefProdDens}) and
taking the continuum limit in both space and time, one arrives at
the evolution equation for the second order product density
\begin{eqnarray}
\nonumber \frac{1}{2}\,\frac{\partial}{\partial
t}\mb(\xi_{1},t)&=& b \int_{\Re^2}
B(\xi_{2})\,\mb(\xi_{1}-\xi_{2},t)\,d\xi_{2}+b\,B(\xi_{1})\,\ma(t)-d\,
\mb(\xi_{1},t) \\  &-&
\dn W(\xi_{1})\,\mb(\xi_{1},t)-\dn \int_{\Re^2}
W(\xi_{2})\,\mc(\xi_{1},\xi_{2},t)\,d\xi_{2},\label{eq:secondProdDensSpat}
\end{eqnarray}
where we see the dependence on the \emph{third} order product
density in the last integral
 \end{appendix}
 
\paragraph{acknowledgements}
M.R acknowledges the support granted by the International Institute for Applied Systems Analysis (IIASA) to participate in the Young Scientist Summer Program where part of this research was conducted during the summer of 2004. M.R. is grateful to Richard Law who suggested to work on this problem and generously shared his time and insights,  Kenneth Lindsay who kindly shared his probabilistic and simulation expertise, and fruitful discussions with Jonathan Dushoff.  The authors are also grateful with Benjamin Bolker, David Murrell and David Grey who helped with  details on the simulations of the point processes and generously provided access to code and manuscripts.  The support of Simon A. Levin and Ioannis G. Kevrekidis is gratefully acknowledged.

\bibliographystyle{plain}

%
%



\end{document}